\documentclass[sigconf,10pt]{acmart}

\usepackage{balance}

\def\BibTeX{{\rm B\kern-.05em{\sc i\kern-.025em b}\kern-.08emT\kern-.1667em\lower.7ex\hbox{E}\kern-.125emX}}

\copyrightyear{2020}
\acmYear{2020}
\setcopyright{acmcopyright}\acmConference[SIGMOD'20]{Proceedings of the 2020 ACM SIGMOD International Conference on Management of Data}{June 14--19, 2020}{Portland, OR, USA}
\acmBooktitle{Proceedings of the 2020 ACM SIGMOD International Conference on Management of Data (SIGMOD'20), June 14--19, 2020, Portland, OR, USA}
\acmPrice{15.00}
\acmDOI{10.1145/3318464.3389754}
\acmISBN{978-1-4503-6735-6/20/06}

\acmSubmissionID{mod0580}

\newcommand{\sys}{\mbox{\textsc{sqlcheck}}\xspace}
\newcommand{\dbdeo}{\mbox{\textsc{dbdeo}}\xspace}

\newcommand{\parser}{\texttt{sqlparse}\xspace}
\newcommand{\detector}{\texttt{ap-detect}\xspace}
\newcommand{\ranker}{\texttt{ap-rank}\xspace}
\newcommand{\fixer}{\texttt{ap-fix}\xspace}

\newcommand{\contextbuilder}{\texttt{ContextBuilder}\xspace}
\newcommand{\apshort}{\texttt{AP}\xspace}
\newcommand{\apsshort}{\texttt{APs}\xspace}
\newcommand{\sql}{SQL\xspace}
\newcommand{\globaleaks}{Globaleaks\xspace}

\newcommand{\gl}{\textsf{GlobaLeaks}\xspace}

\newcommand{\mva}{Multi-Valued Attribute\xspace}
\newcommand{\npk}{No Primary Key\xspace}
\newcommand{\nfk}{No Foreign Key\xspace}
\newcommand{\gpk}{Generic Primary Key\xspace}

\newcommand{\ent}{Enumerated Types\xspace}

\newcommand{\ixu}{Index Underuse\xspace}

\newcommand{\pmr}{Pattern Matching\xspace}

\newcommand{\tmj}{Too Many Joins\xspace}

\newcommand{\appcount}{\texttt{1406}\xspace}

\newcommand{\djangoappcount}{\texttt{15}\xspace}
\newcommand{\djangoapcount}{\texttt{123}\xspace}
\newcommand{\djangomajorapcount}{\texttt{32}\xspace}

\newcommand{\kaggledbcount}{\texttt{31}\xspace}
\newcommand{\kaggleapcount}{\texttt{200}\xspace}

\newcommand{\totalapsupportcount}{\texttt{26}\xspace}

\newcommand{\paperTitle}{SQLCheck: Automated Detection and Diagnosis of SQL Anti-Patterns} 
\newcommand{\paperAuthors}{Prashanth Dintyala,Arpit Narechania,Joy Arulraj}
\newcommand{\paperKeywords}{Anti-Patterns; Database Applications}

\setcopyright{none}
\acmDOI{}
\acmISBN{}
\acmYear{}
\copyrightyear{} 
\acmPrice{}
\acmConference[]{}{}{}

\definecolor{linkcolor}{HTML}{647382}
\definecolor{citecolor}{HTML}{647382} %
\definecolor{urlcolor}{rgb}{0.4,0.2,0.2}
\definecolor{sqlcolor}{HTML}{965d67}
\definecolor{smtcolor}{HTML}{5d968c}
\definecolor{commentcolor}{HTML}{4B0082}

\hypersetup{
    pdfauthor = {\paperAuthors},
    pdftitle = {\paperTitle},
    pdfkeywords = {\paperKeywords},
    bookmarksopen = {true},
    colorlinks=true,
    citecolor={urlcolor},
    linkcolor={linkcolor},
 	urlcolor={citecolor},
    pdfborder={ 0 0 0 }
}

\usepackage{amsmath,amsopn,amssymb}
\usepackage{listings}
\usepackage{fancyvrb}
\usepackage{supertabular,booktabs}
\usepackage{array,underscore,relsize}
\usepackage{enumitem}
\usepackage{balance}
\usepackage{booktabs}
\usepackage{pifont}
\usepackage{multirow}
\usepackage[scaled]{beramono}
\usepackage{tabularx}
\usepackage{graphics}
\usepackage{subfig}

\usepackage[compact]{titlesec}

\newcounter{example}[part]
\newenvironment{example}[1][]{\refstepcounter{example}\par\vspace{4pt}
   \noindent \textbf{\textsc{Example}~\theexample #1:}
   \rmfamily}{\vspace{4pt}}
 
\usepackage{aliascnt}
\usepackage[ruled,linesnumbered,noend]{algorithm2e}

\SetCommentSty{mycommfont}

\usepackage{enumitem}

\SetAlFnt{\footnotesize}
\SetAlCapFnt{\small}
\SetAlCapNameFnt{\small}

\newcommand{\dcircle}[1]{\ding{\numexpr181 + #1}}

\usepackage[capitalize,noabbrev,nameinlink]{cleveref}

\crefname{lstlisting}{listing}{listings}
\Crefname{lstlisting}{Listing}{Listings}

\newcommand{\new}[1]{{\textcolor{black}{#1}}}

\usepackage[compact]{titlesec}
\titlespacing{\section}{0pt}{5pt}{2pt}
\titlespacing{\subsection}{0pt}{3pt}{1pt}
\titlespacing{\subsubsection}{0pt}{3pt}{1pt}

\usepackage{soul}

\usepackage[title]{appendix}

\fvset{fontsize=\scriptsize,xleftmargin=8pt,numbers=left,numbersep=5pt}

\def\Snospace~{\S{}}

\newif\ifdraft\drafttrue
\newif\ifnotes\notestrue
\ifdraft\else\notesfalse\fi

\newcolumntype{R}[1]{>{\raggedleft\let\newline\\\arraybackslash\hspace{0pt}}p{#1}}

\newcommand{\PP}[1]{
\vspace{0.043in}
\noindent{\bf\textsc{#1}}\xspace
}

\newcommand{\squishitemize}{
 \begin{list}{$\bullet$}
  { \setlength{\itemsep}{0pt}
     \setlength{\parsep}{3pt}
     \setlength{\topsep}{3pt}
     \setlength{\partopsep}{0pt}
     \setlength{\leftmargin}{1.95em}
     \setlength{\labelwidth}{1.5em}
     \setlength{\labelsep}{0.5em} } }

\newcounter{Lcount}
\newcommand{\squishlist}{
    \begin{list}{\arabic{Lcount}. }
   { \usecounter{Lcount}
        \setlength{\itemsep}{0pt}
        \setlength{\parsep}{3pt}
        \setlength{\topsep}{3pt}
        \setlength{\partopsep}{0pt}
        \setlength{\leftmargin}{2em}
        \setlength{\labelwidth}{1.5em}
        \setlength{\labelsep}{0.5em} } }

\newcommand{\squishend}{\end{list}}

\usepackage{xstring}

\newcommand{\eg}{\textit{e.g.}\xspace}
\newcommand{\ie}{\textit{i.e.}\xspace}
\newcommand{\etal}{\textit{et al.}\xspace}

\makeatletter
\newcommand\BeraMonottfamily{%
  \def\fvm@Scale{0.85}
  \fontfamily{fvm}\selectfont
}
\makeatother

\lstdefinestyle{SQLStyle}{
language=SQL,
basicstyle=\BeraMonottfamily\footnotesize, 
keywordstyle=\color{sqlcolor}\bfseries,
commentstyle=\color{commentcolor}\ttfamily,
deletekeywords ={Role,ROLE},
literate = {-}{-}1, 
}

\lstdefinestyle{ScriStyle}{
language=SQL,
basicstyle=\BeraMonottfamily\footnotesize, 
keywordstyle=\color{smtcolor}\bfseries,
commentstyle=\color{commentcolor}\ttfamily,
morekeywords={and, or, not},
literate = {-}{-}1, 
}

\lstdefinestyle{PythonStyle}{
language=Python,
basicstyle=\BeraMonottfamily\footnotesize, 
keywordstyle=\color{smtcolor}\bfseries,
commentstyle=\color{commentcolor}\ttfamily,
morekeywords={and, or, not},
literate = {-}{-}1, 
columns=fullflexible,
breaklines=true,
postbreak=\mbox{\textcolor{blue}{$\hookrightarrow$}\space},
}

\definecolor{eclipseStrings}{RGB}{42,0.0,255}
\definecolor{eclipseKeywords}{RGB}{127,0,85}
\colorlet{numb}{magenta!60!black}

\lstdefinelanguage{JSON}{
    basicstyle=\normalfont\ttfamily,
    commentstyle=\color{eclipseStrings}, 
    stringstyle=\color{eclipseKeywords}, 
    showstringspaces=false,
    breaklines=true,
    string=[s]{"}{"},
    comment=[l]{:\ "},
    morecomment=[l]{:"},
    literate=
        *{0}{{{\color{numb}0}}}{1}
         {1}{{{\color{numb}1}}}{1}
         {2}{{{\color{numb}2}}}{1}
         {3}{{{\color{numb}3}}}{1}
         {4}{{{\color{numb}4}}}{1}
         {5}{{{\color{numb}5}}}{1}
         {6}{{{\color{numb}6}}}{1}
         {7}{{{\color{numb}7}}}{1}
         {8}{{{\color{numb}8}}}{1}
         {9}{{{\color{numb}9}}}{1}
}

\lstdefinestyle{JSONStyle}{
language=JSON,
basicstyle=\BeraMonottfamily\footnotesize, 
keywordstyle=\color{smtcolor}\bfseries,
commentstyle=\color{commentcolor}\ttfamily,
morekeywords={and, or, not},
literate = {-}{-}1, 
columns=fullflexible,
breaklines=true,
postbreak=\mbox{\textcolor{blue}{$\hookrightarrow$}\space},
}

\crefname{lstlisting}{listing}{listings}
\Crefname{lstlisting}{Listing}{Listings}

\definecolor{webgreen}{rgb}{0,.5,0}                                                                                                                     
\definecolor{webbrown}{rgb}{.6,0,0}                                                                                                                   
\definecolor{webblue}{rgb}{0,0,.7} 

\usepackage{siunitx}                                                                                                                                  
\usepackage{tikz}

\begin{document}

\fancyhead{}
  
\title{\paperTitle}


\author{Prashanth Dintyala}
\authornote{These authors contributed equally to this work.}
\email{vdintyala3@gatech.edu}
\affiliation{%
  \institution{Georgia Institute of Technology}
}

\author{Arpit Narechania}
\authornotemark[1]
\email{arpitnarechania@gatech.edu}
\affiliation{%
  \institution{Georgia Institute of Technology}
}

\author{Joy Arulraj}
\email{arulraj@gatech.edu}
\affiliation{%
  \institution{Georgia Institute of Technology}
}

%
\renewcommand{\shortauthors}{Dintyala and Narechania, et al.}

\begin{abstract}
The emergence of database-as-a-service platforms has made deploying database
applications easier than before.
Now, developers can quickly create scalable applications.
However, designing performant, maintainable, and accurate applications
is challenging. 
Developers may unknowingly introduce anti-patterns in the application's 
SQL statements.
These anti-patterns are design decisions that are intended to solve a problem,
but often lead to other problems by violating fundamental design principles.

In this paper, we present \sys, a holistic toolchain for automatically
finding and fixing anti-patterns in database applications. 
We introduce techniques for automatically (1) detecting anti-patterns with high
precision and recall, (2) ranking the anti-patterns based on their impact on 
performance, maintainability, and accuracy of applications, and 
(3) suggesting alternative queries and changes to the 
database design to fix these anti-patterns. 
We demonstrate the prevalence of these anti-patterns in a large collection of
queries and databases collected from open-source repositories.
We introduce an anti-pattern detection algorithm that augments query analysis
with data analysis. 
We present a ranking model for characterizing the impact of frequently 
occurring anti-patterns. 
We discuss how \sys suggests fixes for high-impact anti-patterns using
rule-based query refactoring techniques.
%
%
Our experiments demonstrate that \sys enables  developers 
to create more performant, maintainable, and accurate applications.
\end{abstract}


%

\maketitle

\section{Introduction}
\label{sec:introduction}


Two major trends have simplified the design and deployment of data-intensive
applications. 
The first is the spread of data science skills to a larger community of
developers~\cite{data-science,segaran09}.
Data scientists combine rich data sources in applications that process large
amounts of data in real-time.
These applications produce qualitatively better insights in many domains, such
as science, governance, and industry~\cite{davenport12}.
The second trend is the proliferation of database-as-a-service (DBaaS) platforms
in the cloud~\cite{curino11,lomet09}. 
Due to economies of scale, these services enable greater access to database
management systems (DBMSs) that were previously only in reach for large
enterprises.
DBaaS platforms obviate the need for in-house database administrators (DBAs) and
enable data scientists to quickly deploy widely used applications.

\PP{Challenge:}
Designing database applications is, however, non-trivial since applications may
suffer from \textit{anti-patterns}~\cite{karwin10}. 
An anti-pattern (AP) refers to a design decision that is intended to solve a
problem, but that often leads to other problems by violating fundamental design
principles. 
\apsshort in database applications can lead to convoluted logical and physical
database designs, thereby affecting the performance, maintainability,
and accuracy of the application.
The spread of data science skills to a larger community of developers places
increased demand for a toolchain that facilitates application design
without \apsshort since scientists who are experts in other domains
are likely not familiar with these anti-patterns~\cite{data-science,segaran09}.
Furthermore, the proliferation of DBaaS platforms obviates the need for in-house
DBAs who used to assist application developers with finding and fixing 
\apsshort.

Sharma \etal have designed a tool, called \dbdeo, for automatically
detecting \apsshort in database applications~\cite{sharma18}.
They demonstrate the widespread prevalence of \apsshort in production applications.
%
%
%
%
Although their detection algorithm of \dbdeo is effective in uncovering \apsshort, it suffers from three limitations.
First, the static analysis algorithm suffers from low precision and recall.
Second, it does not rank the \apsshort based on their impact.
Third, it does not suggest solutions for fixing them.
Thus, a developer would need to manually confirm the \apsshort detected by
\dbdeo, identify the high-impact \apsshort among them, and fix them.

In this paper, we investigate how to find, rank, and fix \apsshort in database
applications.
We present a toolchain, called \sys, that assists application
developers by:
(1) detecting \apsshort with high precision and recall,
(2) ranking the detected \apsshort based on their impact on performance and
maintainability, 
(3) suggesting fixes for high-impact \apsshort.

The main thrust of our approach is to augment code analysis with data analysis
(\ie examine both queries and data sets of the application) to detect \apsshort with
high precision and recall.
We study the impact of frequently occurring \apsshort on  the performance,
maintainability, and accuracy of the application.
We then use this information to rank the \apsshort based on their estimated impact. 
By targeting frequently occurring \apsshort, we take advantage of our ranking model
trained on data collected from previous deployments without needing to share
sensitive data (\eg data sets).
Lastly, \sys suggests fixes for high-impact \apsshort using rule-based query
refactoring techniques.
The advantage of our approach over \dbdeo is that it reduces the time
that a developer must expend on identifying high-impact \apsshort and fixing
them.

In summary, we make the following contributions:
\squishitemize
\item We illustrate the limitations of the state-of-the-art tools for identifying
\apsshort in database applications and motivate the need for an alternate approach
with higher precision and recall  (\autoref{sec:motivation}).

\item We introduce an AP detection algorithm that augments query analysis with
data analysis (\autoref{sec:detection}).

\item We present a ranking model for characterizing the impact of frequently
occurring \apsshort on the performance, maintainability, and accuracy of the
application (\autoref{sec:ranking}).

\item We discuss how \sys suggests fixes for high-impact \apsshort using
rule-based query refactoring techniques (\autoref{sec:repair}). 

\item We illustrate the efficacy of \sys in finding, ranking, and fixing
\apsshort through an analysis of \new{\appcount open-source database
applications, \djangoappcount Django applications, \texttt{\kaggledbcount} Kaggle databases}, and a user study (\autoref{sec:evaluation}).
\squishend


\section{Motivation \& Background}
\label{sec:motivation}

We illustrate the need for detecting and diagnosing \apsshort through a case
study, then present an overview of the different types of \apsshort and conclude with a discussion on the impact of \apsshort on the application's performance, maintainability, and accuracy.

\subsection{Case Study: GlobaLeaks}
{\label{sec:casestudy}}

We illustrate the problems introduced by \apsshort through a case study of 
\gl, an open-source application for anonymous-whistleblowing
initiatives~\cite{globaleaks}.
The application supports a \textit{multi-tenancy} feature to enable multiple
organizations to accept submissions and direct them to different endpoints
within a single deployment of the application.

\begin{example}
\label{ex:multi-valued-attribute}
~\cref{fig:pattern-1-bad} presents the logical database design of the two tables
associated with this feature\footnote{We distilled the essence of this \apshort
for the sake of presentation.}. 
%
%
Since a given tenant can serve multiple users (\ie, one-to-many relationship),
the application developer decided to store this information as a 
comma-separated list of user identifiers in the \texttt{User\_IDs} column of \texttt{Tenants} table.
While this \textit{multi-valued attribute} design pattern captures the relationship between the two entities without introducing additional tables or columns, it suffers from performance, maintainability, and data integrity
problems.
We illustrate these problems using a set of tasks and associated SQL queries
executed by the application.
\end{example}

\PP{Task \#1:}
The developer is interested in retrieving the tenants that a user is associated
with. We cannot use the equality operator in SQL to solve this task since the 
users are stored in a comma-separated list. Instead, we must employ
\textit{pattern-matching expressions} to search for that user:

\begin{lstlisting}[style=SQLStyle]
/* List the tenants that a user is associated with */
SELECT * FROM Tenants WHERE User_IDs LIKE `[[:<:]]U1[[:>:]]';
\end{lstlisting}

\PP{Task \#2:}
Consider the task of retrieving information about the users served by a
tenant. This query is also computationally expensive since this 
involves joining the comma-separated list of users to matching rows in the
\texttt{Users} table. Joining two tables using an expression prevents the
DBMS from using indexes to accelerate query processing~\cite{graefe93b}.  
Instead, it must scan through both tables, generate a cross product, and
evaluate the regular expression for every combination of rows.

\begin{lstlisting}[style=SQLStyle]
/* Retrieve users served by a tenant */
SELECT * FROM Tenants AS t JOIN Users AS u
ON t.User_IDs LIKE `[[:\<:]]'||u.User_ID||`[[:\>:]]'
WHERE t.Tenant_ID = 'T1';
\end{lstlisting}

%
%

%

\begin{figure}
\centering
\subfloat[Tenants Table]{
\footnotesize
\begin{tabular}{r c c c c}
                   \textbf{Tenant\_ID}  & \textbf{Zone\_ID} & 
                   \textbf{Active} & \textbf{User\_IDs}
    \\
	\midrule
    T1 & Z1 & True & U1 , U2\\
    T2 & Z3 & True & U3 \textbf{;} U4 \\
\end{tabular}

\label{fig:pattern-1-bad-1}
}
\hfill
\subfloat[Users Table]{
\footnotesize
\begin{tabular}{r c c c}
                   \textbf{User\_ID}  & \textbf{Name} &
                   \textbf{Role} & \textbf{Email} 
    \\
	\midrule
    U1  & N1  & R1 & E1 \\
    U2  & N2  & R2 & E2  \\
    U3  & N3  & R3 & E3 \\    
    U4  & N4  & R4 & E4 \\         
\end{tabular}

\label{fig:pattern-1-bad-2}
}
\caption{
\textbf{GlobaLeaks Application} --
List of tables.
}
\label{fig:pattern-1-bad}
\end{figure}
\begin{figure}
\centering
\subfloat[Users Table]{
\footnotesize
\begin{tabular}{r c c c c}
                   \textbf{User\_ID}  & \textbf{Name} &
                   \textbf{Role} & \textbf{Email}                   
    \\
	\midrule

    U1  & N1 & R1 &  E1 \\
\end{tabular}

\label{fig:pattern-1-good-2}
}
\hfill
\subfloat[Tenants Table]{
\footnotesize
\begin{tabular}{r c c }
                   \textbf{Tenant\_ID}  & \textbf{Zone\_ID} &  \textbf{Active} 
    \\
	\midrule

    T1  & Z1 & True \\
    T2  & Z2 & True \\
\end{tabular}

\label{fig:pattern-1-good-1}
}
\hfill
\subfloat[Hosting Table]{
\footnotesize
\begin{tabular}{c c} 
				 \textbf{Tenant\_ID}  & \textbf{User\_ID}
    \\
    \hline
    T1 & U1  \\
    T1 & U2  \\
    T2 & U3 \\
    T2 & U4 \\
\end{tabular}

\label{fig:pattern-1-good-3}
}
\caption{
\textbf{Refactored GlobaLeaks Application} --
List of tables.
}
\label{fig:pattern-1-good}
\end{figure}

\PP{Data Integrity Problems:}
Another major limitation of this approach is that the developer implicitly
assumes that users will be stored as a list of strings separated by a comma. 
This implicit assumption might later be violated by a developer 
entering users separated by another delimiter, such as a semi-colon (\eg
``U6\textbf{;} U7''). 
This is feasible since the DBMS is not explicitly enforcing that the string
should be separated by a particular character.
Given this new data, the developer must update all the queries operating on the
\texttt{User\_IDs} column to handle the usage of multiple separator
characters.
Furthermore, it is not feasible for the DBMS to enforce a referential integrity
constraint between these columns: (1) \texttt{User\_IDs} in \texttt{Tenants},
and (2) \texttt{User\_ID} in \texttt{Users}. This is because the former column
encodes the comma-separated list as a string. 
So, it is possible for a user in the former column to not have a corresponding
tuple in the latter column.

\subsubsection{Solution: Intersection Table } 

 We can eliminate this \apshort  by creating an additional \textit{intersection
table} to encode the many-to-many relationship between tenants and users
\footnote{That is, each user may be associated with multiple tenants, and
likewise each tenant may serve multiple users.}.
This table references the \texttt{Tenant} and \texttt{User} tables.
In~\cref{fig:pattern-1-good}, the \texttt{Hosting} table implements this
relationship between the two referenced tables.

\begin{lstlisting}[style=SQLStyle]  
/* Create an intersection table */
CREATE TABLE Hosting (
    User_ID VARCHAR(10) REFERENCES User(User_ID), 
    Tenant_ID VARCHAR(10) REFERENCES Tenants(Tenant_ID),
    PRIMARY KEY (User_ID, Tenant_ID)
);
/* Drop redundant column */
ALTER TABLE Tenants DROP COLUMN User_IDs;
\end{lstlisting}

We will next illustrate how this \apshort-free logical design enables 
simpler queries for all of the tasks.

\PP{Tasks \#1 and \#2:}
It is straightforward to join the \texttt{Tenants} and \texttt{Users}
tables with the \texttt{Hosting} table to solve the first two tasks.
These queries are easy to write for developers and easy to optimize for
DBMSs. The DBMS can now use an index on \texttt{User\_IDs} to efficiently
execute the join instead of matching regular expressions.



\begin{lstlisting}[style=SQLStyle]
/* List the tenants that a user is associated with */
SELECT * FROM Hosting as H JOIN Tenants as T
ON H.Tenant_ID == T.Tenant_ID WHERE H.User_ID = 'U1';
/* Retrieve information about users served by tenant */
SELECT * FROM Hosting as H JOIN Tenants as T
ON H.User_ID == T.User_ID WHERE H.Tenant_ID = 'T1';
\end{lstlisting}

%


%

\PP{Data Integrity Problems:}
The developer can delegate the task of ensuring data integrity to the DBMS by
specifying the appropriate foreign key constraints.
The DBMS will enforce these constraints when data is ingested or updated. 


\subsection{Classification of Anti-Patterns}
\label{sec:motivation::classification}

\begin{table*}
    \centering
    \footnotesize
    
\newcolumntype{Y}{>{\centering\arraybackslash}X}
\begin{tabularx}{\textwidth}{@{}llllllll@{}}
\toprule
\textbf{Category} & 
\textbf{Anti-Pattern Name} &
\textbf{Description} &
\textbf{P} &  
\textbf{M} & 
\textbf{DA} &
\textbf{DI} & 
\textbf{A}
\\
\midrule
\multirow{7}{*}{Logical Design \apsshort}
& Multi-Valued Attribute 
& \multicolumn{1}{p{8cm}}{Storing list of values
in a delimiter-separated list violating 1-NF.} &
$\checkmark$  & $\checkmark$ & $\checkmark^{(\downarrow)}$ & $\checkmark$ & $\checkmark$
\\
& No Primary Key 
& \multicolumn{1}{p{8cm}}{Lack of data integrity constraints.} &
$\checkmark$  & $\checkmark$ & $\checkmark^{(\uparrow)}$ & $\checkmark$ & -
\\
& No Foreign Key 
& \multicolumn{1}{p{8cm}}{Lack of referential integrity constraints.} &
$\checkmark$  & $\checkmark$ & - &  $\checkmark$ & - 
\\
& Generic Primary Key 
& \multicolumn{1}{p{8cm}}{Creating a generic primary key column (\eg, \texttt{id}) for each table.} & 
- &  $\checkmark$ & - &  - & - \\
& Data In Metadata 
& \multicolumn{1}{p{8cm}}{\raggedright 
Hard-coding application logic in table's meta-data.
} &
$\checkmark$  & $\checkmark$ & $\checkmark^{(\downarrow)}$ & $\checkmark$ & $\checkmark$ 
\\
& Adjacency List 
& \multicolumn{1}{p{8cm}}{Foreign key constraint referring to an attribute in the same table.} & $\checkmark$ & - & -  & - & - \\
& God Table 
& \multicolumn{1}{p{8cm}}{Number of attributes defined in the table cross a
threshold (\eg, 10)} & $\checkmark$ &  $\checkmark$  & -  & - & - \\

\midrule
\multirow{6}{*}{Physical Design \apsshort}
& Rounding Errors 
& \multicolumn{1}{p{8cm}}{Storing fractional data using
a type with finite precision (\eg, \texttt{FLOAT}).} &
- & - & -  & - &  $\checkmark$ \\
& Enumerated Types
& \multicolumn{1}{p{8cm}}{Using \texttt{enum} to
constrain the domain of a column.} &
$\checkmark$  & $\checkmark$ & $\checkmark^{(\downarrow)}$ & - & -  \\
& External Data Storage 
& \multicolumn{1}{p{8cm}}{Storing file paths 
instead of actual file content in database. } &
- & $\checkmark$  & - & $\checkmark$ & $\checkmark$ \\
& Index Overuse 
& \multicolumn{1}{p{8cm}}{Creating too many 
infrequently-used indexes.} &
$\checkmark$  & $\checkmark$ & $\checkmark^{(\downarrow)}$ & - & - \\
& Index Underuse 
& \multicolumn{1}{p{8cm}}{Lack of performance-critical
indexes.} &
$\checkmark$  & $\checkmark$ & $\checkmark^{(\uparrow)}$ & - & - \\
& Clone Table
 & \multicolumn{1}{p{8cm}}{Multiple tables matching the pattern <TableName>_N} &
 $\checkmark$ & $\checkmark$ & - & $\checkmark$ & $\checkmark$ \\

\midrule
\multirow{7}{*}{Query \apsshort}
& Column Wildcard Usage 
& \multicolumn{1}{p{8cm}}{Selecting all
attributes from a table using wildcards to reduce typing.} &
$\checkmark$ & - & -  & - &  $\checkmark$\\
& Concatenate Nulls
& \multicolumn{1}{p{8cm}}{Concatenating columns that
might contain \texttt{NULL} values using \texttt{||}.} &
- & - & -  & - &  $\checkmark$ \\
& Ordering by \texttt{RAND} 
& \multicolumn{1}{p{8cm}}{\raggedright Using
\texttt{RAND} function for random sampling or shuffling.} &
$\checkmark$ & - & -  & - & - \\
& Pattern Matching 
& \multicolumn{1}{p{8cm}}{Using regular expressions for
pattern matching complex strings.} &
$\checkmark$ & - & -  & - & - \\
& Implicit Columns 
& \multicolumn{1}{p{8cm}}{
Not explicitly specifying column names in data modification operations.} &
-  & $\checkmark$ & - & $\checkmark$ & -  \\
& \texttt{DISTINCT} and \texttt{JOIN} 
& \multicolumn{1}{p{8cm}}{Using \texttt{DISTINCT} to remove duplicate values
generated by a \texttt{JOIN}.} &
$\checkmark$ &  $\checkmark$ & - & - & - \\
& Too Many Joins 
& \multicolumn{1}{p{8cm}}{Number of JOINs cross a threshold.} &
$\checkmark$ & - & -  & - & - \\


\midrule
\multirow{6}{*}{\new{Data \apsshort}}
& \new{Missing Timezone}
& \multicolumn{1}{p{8cm}}{\new{Date-time fields stored without timezone.}} 
& - & - & -  & - & $\checkmark$ \\
& \new{Incorrect Data Type}
& \multicolumn{1}{p{8cm}}{\new{Actual data does not conform to expected data
type.}} & $\checkmark$ & - & $\checkmark^{(\downarrow)}$  & - & - \\
& \new{Denormalized Table}
& \multicolumn{1}{p{8cm}}{\new{Duplication of values.}} 
& $\checkmark$ & - & $\checkmark^{(\downarrow)}$  & - & - \\
& \new{Information Duplication}
& \multicolumn{1}{p{8cm}}{\new{Derived columns (\eg, age from date of birth).}}
& - & $\checkmark$ & -  & $\checkmark$ & $\checkmark$ \\
& \new{Redundant Column}
& \multicolumn{1}{p{8cm}}{\new{Column with NULLS or same value (\eg,
en-us)}} & - & - & $\checkmark^{(\downarrow)}$  & - & - \\
& \new{No Domain Constraint}
& \multicolumn{1}{p{8cm}}{\new{All values should belong to particular range
(\eg, rating)}} & - & $\checkmark$ & $\checkmark^{(\downarrow)}$  & $\checkmark$
& - \\

\bottomrule
\end{tabularx}

    \caption{\textbf{List of Anti-Patterns:} A catalog of \apsshort 
    based on best practices for database application design~\protect\cite{karwin10,stackoverflow,c2,sharma18}.
    They fall under four categories: (1) logical design \apsshort, (2) physical design \apsshort,  (3) query \apsshort, and (4) data \apsshort.
    For each \apshort we illustrate its impact on five metrics:
    (1) Performance (P), (2) Maintainability (M),   
    (3) Data Amplification (DA), (4)  Data Integrity (DI), and 
    (5) Accuracy (A).   
    $\checkmark$ represents that the given \apshort affects that metric. 
    $\uparrow$ and $\downarrow$ refer to increase and decrease in data 
    amplification, respectively, when that \apshort is fixed. 
    }
    
    \label{tab:ap-list}
\end{table*}

We compiled a catalog of \apsshort based on several resources 
that discuss best practices for schema design and query
structuring~\cite{karwin10,stackoverflow,c2,sharma18}.
\cref{tab:ap-list} lists the \apsshort that \sys targets.
These \apsshort fall under \new{four} categories:

\PP{\dcircle{1} Logical Design \apsshort:}
This category of \apsshort arises from violating logical design principles that
suggest the best way to organize and interconnect data.
%
It includes the \textit{multi-valued attribute} \apshort covered
in~\autoref{sec:casestudy}.
%
%
%
The \textit{adjacency list} \apshort also falls under this category.
It refers to references between two attributes within the same table. 
Such a logical design is used to model hierarchical structures (\eg
employee-manager relationship).
With this representation, however, it is not trivial to handle common tasks 
such as retrieving the employees of a manager up to a certain depth and
maintaining the integrity of the relationships when a manager is removed.

\PP{\dcircle{2} Physical Design \apsshort:}
The next category of \apsshort is associated with efficiently implementing the 
logical design using the features of a DBMS. This includes \textit{rounding
errors} and \textit{enumerated types} \apsshort.
%
The rounding errors \apshort arises when a scientist uses a type with finite
precision, such as \texttt{FLOAT} to store fractional data. 
This may introduce accuracy problems in queries that calculate aggregates.
The enumerated types \apshort occurs when a scientist restricts a column's
values by specifying the fixed set of values it can take while defining the
table.
However, this \apshort makes it challenging to 
add, remove, or modify permitted values later and reduces the application's
portability\footnote{\texttt{ENUM} data type is a proprietary
feature in the MySQL DBMS.}.
%
%

\PP{\dcircle{3} Query \apsshort:}
Query \apsshort arise from violating practices that suggest the best way to retrieve
and manipulate data using \sql. This includes \textit{\texttt{NULL}
usage}
and \textit{column wildcard usage} \apsshort. 
Developers are often caught off-guard by the behavior of \texttt{NULL} in
\sql. Unlike in most programming languages, \sql treats \texttt{NULL} as a special value, 
different from zero, false, or an empty string. This results in
counter-intuitive query results and introduces accuracy problems. 
The latter \apshort arises when a developer uses wildcards (\texttt{SELECT *}) 
to retrieve all the columns in a table with less typing. This AP, 
however breaks applications on refactoring.


\PP{\dcircle{4} \new{Data \apsshort:}}
\new{
Data \apsshort are a subset of \apsshort that \sys detects by analysing the
data (as opposed to queries).
This includes the \textit{Incorrect Data Type} and \textit{Information
Duplication} \apsshort.
The former \apshort arises due to data type mismatches (\eg, storing a numerical
field in a TEXT column). 
This negatively impacts performance and leads to data amplification.
The latter \apshort occurs when a column contains data derived from another
column in the same table (\eg, storing \texttt{age} based on \texttt{date
of birth}).
While this accelerates query processing, it reduces maintainability and leads
to data amplification.
}
%

\subsection{Impact of Anti-Patterns}
\label{sec:motivation::impact}
\apsshort in database applications lead to convoluted logical and physical
database designs, thereby affecting the performance, maintainability, 
and accuracy of the application.

\PP{1. Performance:}
An application's performance is often measured in terms of throughput
(\eg the number of requests that can be processed per second)
and latency (\eg the time that it takes for the system to respond to a
request)~\cite{ramakrishnan02}. Optimizing these metrics is
important because they determine how quickly an application can process data
and how quickly a user can leverage the application to make new decisions.

\begin{figure}[t]
\centering
\subfloat[Task \#1]{
\includegraphics[width=0.13\textwidth]{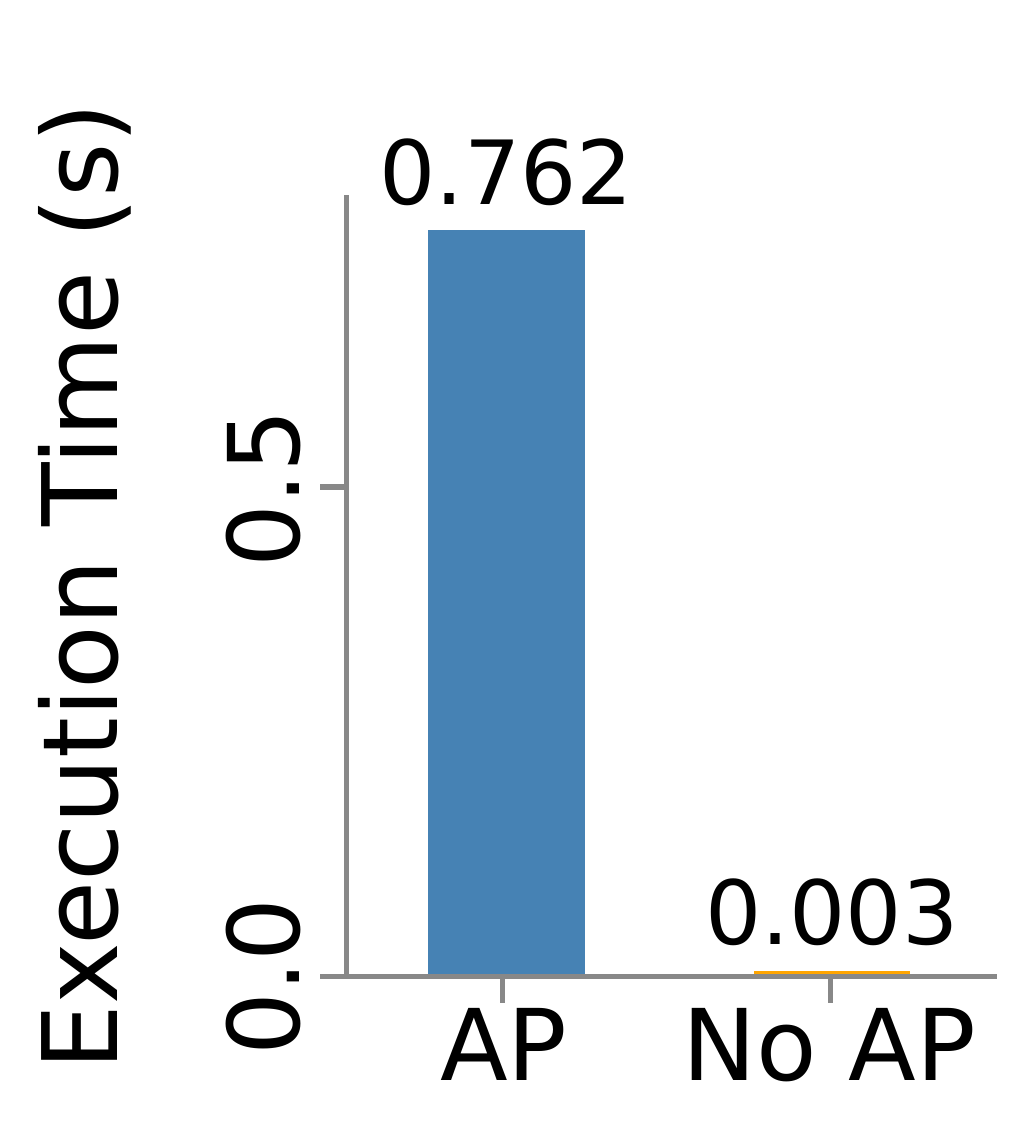}
\label{fig:mva-delete}
}
\subfloat[Task \#2]{
\includegraphics[width=0.13\textwidth]{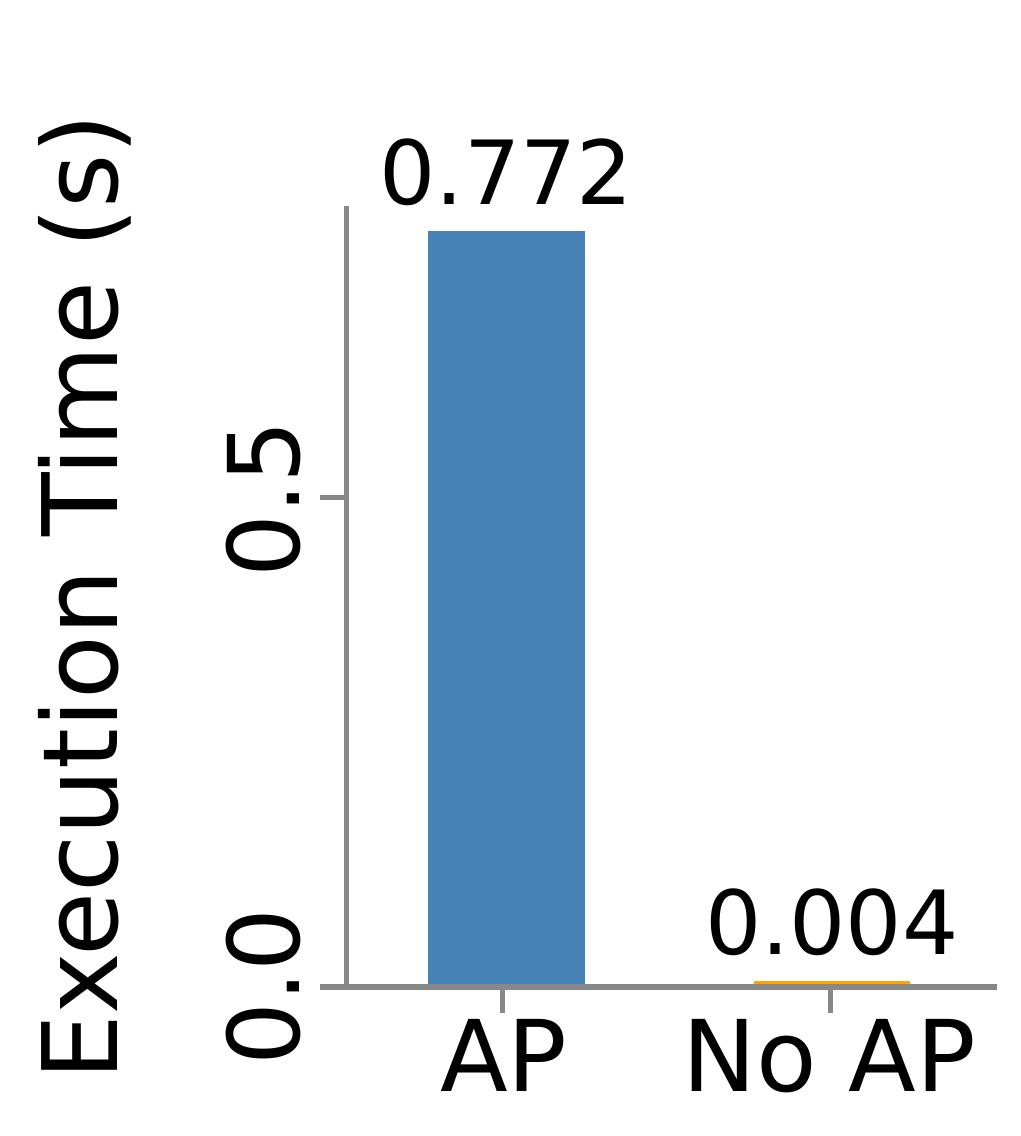}
\label{fig:mva-insert}
}
\subfloat[Task \#3]{
\includegraphics[width=0.13\textwidth]{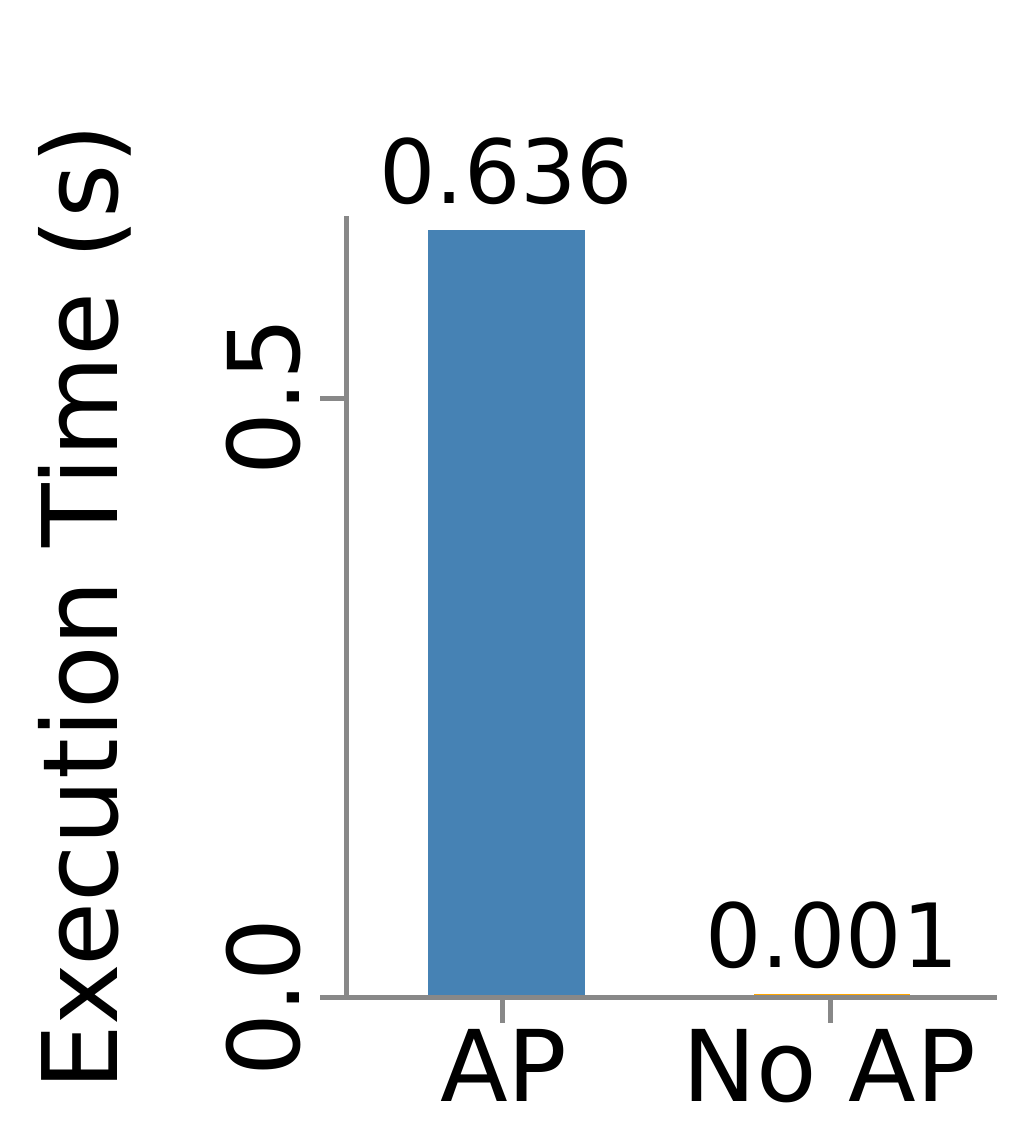}
\label{fig:mva-select}
}
\caption{
\textbf{Multi-Valued Attribute \apshort:} Performance impact of the 
multi-valued attribute \apshort on the above-mentioned tasks.
}
\label{fig:mva-box}
\end{figure}

Consider the tasks presented in~\autoref{sec:motivation::impact}.
We measured the impact of the \textit{multi-valued attribute} \apsshort on 
the time taken to execute these tasks.
~\cref{fig:mva-box} presents the results of this experiment.
We defer the discussion of the experimental setup
to~\autoref{evaluation::ranking-repair}.
When we remove this \apshort, the queries associated with these tasks 
accelerated by 636$\times$, 256$\times$, and 193$\times$
respectively.
These results illustrate the importance of fixing \apsshort.

\PP{2. Maintainability:}
The maintainability of an application represents the ease with which the
application's design and component queries can be modified to adapt to a
changed environment, improve performance or other metrics, 
or  correct faults~\cite{ieee90}. 
Maintainable applications allow developers to quickly and easily add new
features, fix bugs in existing features, and increase performance.

\PP{3. Accuracy:}
An application's accuracy is measured in terms of the discrepancy between the
data stored by a user and that returned by the application. 
For example, an application that stores fractional numeric data using the
\texttt{FLOAT} type in \sql can fail to return certain tuples due to slight
discrepancies in their values~\cite{karwin10}.


Given the impact of \apsshort, we next present an overview of a toolchain that assists developers in eliminating them.
%
%


\section{System Overview}
{\label{sec:thetool}}

\begin{figure}[!t]
  \centering
  \includegraphics[width=0.48\textwidth]{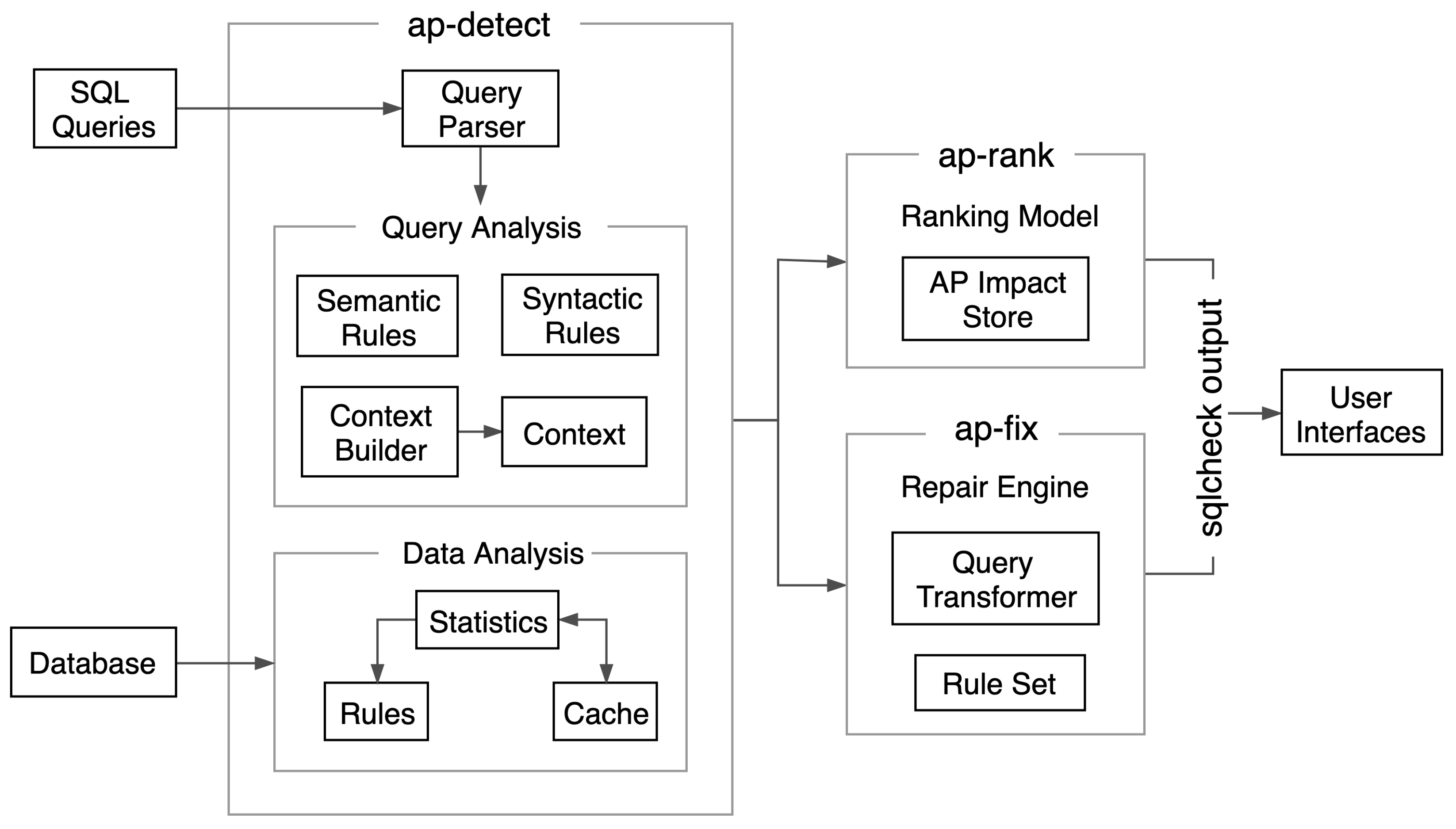}
  \caption{\textbf{Architecture of \sys}:
    It takes in a SQL query and database (optional), and produces a
    ranked list of \apsshort and associated fixes.
    Internally, \sys leverages query and data analysis to detect the 
    \apsshort. It then uses a ranking model and query repair engine to 
    generate the desired fixes.
  }
  \label{fig:architecture-high-level}
\end{figure}

\sys is geared towards automatically finding, ranking, and fixing 
\apsshort in database applications.
Application developers can leverage \sys to create more performant,
maintainable, and accurate database applications.
\sys contains three components for: (1) detecting \apsshort with high
precision and recall, (2) ranking the detected \apsshort based on their impact, 
and (3) suggesting fixes to application developers for high-impact \apsshort.

\PP{Workflow:}
We envision that application developers will use \sys in the following
manner.
A developer will deploy \sys on their local machine and connect it to 
the target application (\ie, queries and data sets).
\dcircle{1} The first component, \detector, will then perform static analysis of
the queries to detect \apsshort. 
To increase precision and recall, \detector will profile the application's 
data and meta-data (\autoref{sec:detection}).
\dcircle{2} Next, \ranker will examine the \apsshort detected by \detector in
the target application and rank them based on their estimated impact
(\autoref{sec:ranking}).
\dcircle{3} The third tool, \fixer, will suggest fixes for the high-impact
\apsshort identified by \ranker using rule-based query transformations
(\autoref{sec:repair}).
\dcircle{4} Lastly, \sys will optionally upload the \apsshort detected in
the application to an online \apshort repository with the permission of the
developer. As new performance data is collected over time, \ranker will retrain
its ranking model to improve the quality of its decisions.

\section{Finding Anti-Patterns}
{\label{sec:detection}}

\begin{algorithm}[t]
\SetKwFunction{staticDetection}{\textsc{Query-Rules}}
\SetKwFunction{dynamicDetection}{\textsc{Data-Rules}}
\SetKwFunction{buildContext}{\textsc{Context-Builder}}
\SetKwFunction{columns}{\textsc{Metadata}}
\SetKwFunction{queryAnalyser}{\textsc{Query-Analyser}}
\SetKwFunction{dataAnalyser}{\textsc{Data-Analyser}}
\SetKwFunction{repair}{\textsc{AP-Fix}}
\SetKwFunction{rank}{\textsc{AP-Rank}}
\SetKwFunction{stats}{\textsc{Stats}}
\SetAlgoLined
\SetKwInOut{Input}{input} 
\SetKwInOut{Output}{output}
\caption{\sys Algorithm}
\label{algo:detect}
\Input{application queries $\mathcal{Q}$, database $\mathcal{D}$}
\Output{detected \apsshort and associated fixes}

    \tcp{Extract context from queries}    
    query context $\mathcal{QC}$ $\gets$ $\{$ $\}$\\
    \For{query $q$ $in$ $\mathcal{Q}$} {
    \label{algo-line:parse-object-start}
        $\mathcal{QC}$.append(\queryAnalyser{$q$})
        \label{algo-line:parse-object-end} }
         
    \tcp{Extract context from data}  
    data context $\mathcal{DC}$ $\gets$ $\{$ $\}$\\
    \For{table $t$ $in$
    $\mathcal{D}$.tables}{
        \label{algo-line:connection-extraction-start} 
        $\mathcal{DC}$.append(\dataAnalyser{$t$})        
        \label{algo-line:connection-extraction-end}
    }
    
    context $\mathcal{C}$ = \buildContext{$\mathcal{QC}$, $\mathcal{DC}$} \\

    \tcp{Detect anti-patterns}
    anti-patterns $\mathcal{P}$ $\gets$ $\{$ $\}$ \\
    \For{query $q$ $in$ $\mathcal{Q}$}{
        $\mathcal{P}$.append(\staticDetection{$q$, $\mathcal{C}$})
    }
    
	\For{table $t$ $in$ $\mathcal{D}$.tables}{
   		$\mathcal{P}$.append(\dynamicDetection{$t$, $\mathcal{C}$})\\
    }

    ranked anti-patterns $\mathcal{R}$ $\gets$ \rank{$\mathcal{P}$} \\
    \label{algo-line:rank}
 \textbf{return} \repair{$\mathcal{R}$}

\end{algorithm}

In this section, we present the techniques used by \detector for identifying
\apsshort in a given application. 
We begin with an overview of the \sys algorithm.

\PP{Overview:}
As shown in~\cref{algo:detect}, \detector takes a list of \sql queries executed
by the application and a connection to the database server as input.
It constructs an \textit{application context} using the given inputs.
First, it uses a \textit{query analyser} to extract context from the queries
(\eg, column names, table names, predicates, constraints, and indexes).
It tailors the analysis based on the type of \sql statement.
We present the query analyser in~\autoref{sec:finding::query}.
Next, it uses a \textit{data analyser} to extract context from the tables 
in the database (\eg, data distribution and format of each column).
We present the data analyser in~\autoref{sec:finding::data}.
Based on the constructed application context, \detector uses a set of rules for
identifying \apsshort in the given queries\footnote{This includes both data
definition language (DDL) and data manipulation language (DML)
commands~\cite{sql92}.}.
\new{
These rules are general-purpose functions that leverage the overall context of the
application:
(1) queries, (2) data, and (3) meta-data.
The {\apsshort} identified using such query and data analyses are
then ordered by {\ranker}, which will be covered in~\autoref{sec:ranking}.
}

\subsection{Query Analysis}
\label{sec:finding::query}

\detector begins by analysing the \sql queries $\mathcal{Q}$ executed by the
target database application.
It detects \apsshort in the given queries in two phases: 
(1) intra-query detection and (2) \new{inter-query}  detection.
During the first phase, it identifies \apsshort in each query $q$ in
$\mathcal{Q}$.
During the second phase, it leverages the entire context of the application
(\ie, other queries in $\mathcal{Q}$ and logical design of the database $D$) to
detect more complex \apsshort.
We next discuss these techniques in detail.


\PP{\dcircle{1} Intra-Query Detection:}
\detector applies a set of \textit{query rules} on the given query $q$.
\new{
Each query rule consists of a general-purpose function to identify the 
existence of the target {\apshort} in $q$.
Rules can range from a simple, pattern-matching function that uses a set of
regular expressions to complex functions that leverage the inferred context of
the application.
In order to support diverse {\sql} dialects,
}
%
%
\detector leverages a non-validating \sql parser, called \parser
\cite{albrecht18}, to process the \sql statement.
This parser supports multiple dialects by virtue of its non-validating
parsing logic.
However, unlike a typical DBMS parser~\cite{pgparser}, it does not generate a
semantically-rich parse tree.
%
%
We address this limitation by annotating the parse tree returned by \parser.
\detector and \fixer use this annotated parse tree for finding \apsshort and
suggesting fixes, respectively.
The tree-structured representation (as opposed to the raw \sql string) 
allows recursive application of rules and improves the extensibility of the
rule system.


\begin{example}
\label{ex:implicit-column-usage}
Consider the following \sql statement for inserting a record into the
\texttt{TENANT} table:
\begin{lstlisting}[style=SQLStyle] 
INSERT INTO Tenant VALUES ('T1', 'Z1', True, 'U1,U2');
\end{lstlisting}
This DML statement fails to correctly function when the schema of
\texttt{TENANT} table evolves.
For instance, if we drop the \texttt{User\_IDs} column and add a new column
termed \texttt{Description}, it would incorrectly insert values into the table.
This \textit{implicit column usage} \apshort also reduces the maintainability of
the application.
This is because explicitly specifying the column names improves the readability
of the query for another developer who is trying to infer the values being
inserted into each column.
\detector identifies this \apshort by checking whether the column names are
present in the \texttt{INSERT} statement.
An intra-query detection rule is sufficient to detect the existence of this
\apshort.
However, to suggest a fixed \texttt{INSERT} statement, \fixer needs the
application's context (\ie, the schema of the \texttt{TENANT} table).
\end{example}

\PP{\dcircle{2} Inter-query detection:} 
The intra-query detection technique suffers from low precision and recall
as it does not leverage the relationship between queries in the application
which is critical for detecting complex \apsshort.

\begin{example}
\label{ex:no-foreign-key}
Consider the \textit{No Foreign Key} \apshort in \gl\cite{globaleaks}.
There are two tables: \texttt{TENANT} and \texttt{QUESTIONNAIRE}.
The \texttt{Tenant_ID} column should connect these two tables.
However, the DDL statement of the \texttt{QUESTIONNAIRE} table 
does not define this foreign key relationship.
Since the intra-query detection technique applies the rules to each query
independently, it is unable to detect this \apshort by separately examining the
two DDL statements.
\detector can detect the missing foreign key only if it considers both
DDL statements along with the \texttt{JOIN} condition in the \texttt{SELECT}
query as follows.

\begin{lstlisting}[style=SQLStyle]
/* Tenant table */
CREATE TABLE Tenant(Tenant_ID INTEGER PRIMARY KEY,
Zone_ID VARCHAR(30) NOT NULL, Active BOOLEAN);
/* Questionnaire table */
CREATE TABLE Questionnaire (Questionnaire_ID UUID PRIMARY KEY,
Tenant_ID INTEGER, Name VARCHAR(30), Editable BOOLEAN);
/* Select query */
SELECT q.Name, q.Editable, t.Active
FROM   Questionnaire q JOIN Tenant T 
ON T.Tenant_ID = Q.Tenant_ID WHERE q.Editable = true; 
\end{lstlisting}
\end{example}
We address the limitations of the intra-query detection technique
by constructing the application's \textit{context}.
The context contains two components: (1) the schema and (2) the queries
associated with the application.
The \apshort detection rules utilize the context to resolve cases where the
presence or absence of an \apshort cannot be determined with high precision by only
looking at a given query.

\begin{algorithm}[t]
\SetKwFunction{rulesFor}{\textsc{Rules}\textsc{For}\textsc{Query}}
\SetKwFunction{extractMoreInfo}{\textsc{Relevant}\textsc{Rules}}
\SetAlgoLined
\SetKwInOut{Input}{input} 
\SetKwInOut{Output}{output}
\caption{Detecting Anti-Patterns via Query Analysis}
\label{algorithm:query-analysis}
\Input{application query $q$, context $\mathcal{C}$}
\Output{detected \apsshort}
    
    anti-patterns $\mathcal{P}$ $\gets$ $\{$ $\}$ \\
    \tcp{anti-pattern detection rules based on type of query}
    rules $\mathcal{R}$ $\gets$ \rulesFor{$q$} \\
    \For{rule $r$ \textbf{in} $\mathcal{R}$}{
        anti-pattern $p$ $\gets$ $r$($q$, $\mathcal{C}$) \\
        \tcp{use context to identify relevant contextual rules}
        contextual rules $\mathcal{F}$ $\gets$ \extractMoreInfo{$p$, $q$,
        $\mathcal{C}$}
        \\
        
        \tcp{use contextual rules to reduce false positives and negatives}
        \uIf{$\mathcal{F}$($\mathcal{C}$, $q$, $p$)}{
            $\mathcal{P}$.append($p$)
        }
    
    } 
 \textbf{return} $\mathcal{P}$ 
\end{algorithm}

\cref{algorithm:query-analysis} presents the algorithm used for detecting
\apsshort by analyzing queries.
The \contextbuilder constructs the context using the analysed queries and the
database.
The context exports a queryable interface for applying contextual rules on the
queries, schema, and other application-specific metadata.
If the database is not available, the \contextbuilder leverages the DDL
statements to construct the context.
Given the context, \detector first applies a set of \apshort detection rules
based on the type of the query.
It then uses the context to identify the relevant context specific rules that are
subsequently applied to reduce false positives and negatives.

\PP{Limitation:}
The inter-query detection technique also suffers from false
positives and negatives.
Consider the multi-valued attribute \apshort discussed
in~\cref{ex:multi-valued-attribute}.
\detector uses a pattern-matching rule (\ie, regular expression) for detecting 
this \apshort in \texttt{SELECT} queries containing string processing tricks.
However, this rule can result in: (1) false negatives if the delimiter-separated
strings are handled externally in the application code, and (2) false positives
if the delimiter is used for an alternate purpose in application (\eg,
\texttt{ADDRESS} attribute).
Thus, it is not feasible to identify this \apsshort with high precision and
recall by only examining the \sql queries.
We next discuss how \detector extracts and utilizes context from the tables in
the database to overcome this limitation.

\subsection{Data Analysis}
\label{sec:finding::data}

\detector leverages data analysis to improve the precision and recall of
anti-pattern detection.
It uses a \textit{data analyzer} to profile the contents of the database 
used by the application and uses it to augment the context described
in~\autoref{sec:finding::query}.
This information is also used for retrieving the relevant contextual rules while
detecting \apsshort via query analysis.

For example, in case of the multi-valued attribute \apshort
(\cref{ex:multi-valued-attribute}), \detector uses a \textit{data rule} that
checks whether a particular column contains delimiter-separated strings.
It first checks the data type of the column. 
If the column is a \texttt{VARCHAR} or \texttt{TEXT} field, it samples the
columnar data and checks whether that contains delimiter-separated strings.
In case of the \texttt{TENANT} table, as shown in~\cref{fig:pattern-1-bad-1},
the \texttt{User\_IDs} column contains comma-separated strings.
Even if the query rules are unable to detect this \apshort, 
the data rule will correctly flag this column as suffering from the MVA
\apshort.

\begin{algorithm}[t]
\SetKwFunction{sample}{\textsc{Sample}}
\SetAlgoLined
\SetKwInOut{Input}{input} 
\SetKwInOut{Output}{output}
\caption{Detecting Anti-Patterns via Data Analysis}
\Input{context $\mathcal{C}$, database $\mathcal{D}$}
\Output{detected \apsshort}

    anti-patterns $\mathcal{P}$ $\gets$ $\{$ $\}$ \\
    \For{rule $d$ \textbf{in} data rules $\mathcal{D}$}{
		\For{table $t$ $in$ $\mathcal{D}$.tables}{
            \tcp{sample tuples from the table}
            sampled tuples $s$ = \sample{$t$}
            
			\tcp{use data rules to reduce false positives and negatives}
            \uIf{r(C[t], s)}{
	           $\mathcal{P}$.append($p$)
            }
        }
    } 

 \textbf{return} $\mathcal{P}$ 

\end{algorithm}

The data analyzer first scans the database to collect: (1) the schemata of the
component tables, and (2) the distribution of the data in the component columns
(\eg, unique values, mean, median, etc.).
It then collects samples from each table in the examined database.
\detector applies a set of rules for determining the existence of \apsshort in
the sampled data.
If one of these rules is activated, then \detector appends the associated
anti-pattern to the list of \apsshort sent to \ranker.
\begin{example}
\label{ex:enumerated-types}
Consider the following \apshort in \gl.
The \texttt{Role} column in the \texttt{USER} table represents the roles
assumed by the users.
The developer chose to encode this data as a \texttt{STRING} field with a
constraint on the field's domain.

\begin{lstlisting}[style=SQLStyle]
ALTER TABLE User ADD CONSTRAINT User_Role_Check 
CHECK (ROLE IN ('R1', 'R2', 'R3'));
\end{lstlisting}

We refer to this as the \ent \apshort.
In this case, the data analyzer extracts the type information of the
\texttt{Role} column and notices that it as a \texttt{STRING} field.
It then samples the data in the column.
\detector uses the context to compute the ratio of distinct values to the number of tuples.
If this ratio is greater than a given threshold, it detects this \apshort.
\end{example}

Since data analysis is computationally expensive (\eg, sampling), \detector
reuses the constructed context across several checks.
The data analyzer periodically refreshes the context over time.
It also refreshes the context whenever the schema evolves.
\detector allows the developer to configure the tuple sampling frequency and 
the thresholds associated with activating data rules.

\PP{\new{Rule Complexity:}}
\new{
\detector supports complex, general-purpose rules that leverage the overall
context of the application.
\cref{ex:index-overuse} illustrates the complexity of rules.
}

\begin{example}
\label{ex:index-overuse}
\new{
The \textit{Index Overuse} \apshort is associated with the creation of too many
infrequently-used indexes.
For instance, consider these three indexes in the \texttt{TENANT} table.
}
\begin{lstlisting}[style=SQLStyle] 
CREATE INDEX idx_zone_actv (Zone_ID, Active); /* Index 1 */
CREATE INDEX idx_zone (Zone_ID); /* Index 2 */
CREATE INDEX idx_actv (Active); /* Index 3 */
/* Queries (Workload 1) */
SELECT Tenant_ID FROM Tenant WHERE Zone_ID = 'Z1' 
 AND Active = 'True';
SELECT Tenant_ID FROM Tenant WHERE Tenant_ID = 'T1'
AND Active = 'True';
/* Queries (Workload 2) */
SELECT Tenant_ID FROM Tenant WHERE Zone_ID = 'Z1';
SELECT Tenant_ID FROM Tenant WHERE Active = 'True';
\end{lstlisting}
Depending on the workload, \detector marks different set of indexes as
potentially exhibiting this \apshort.
It leverages the context to determine the list of constructed indexes.
For the first workload, it marks the second and third indexes as 
redundant since these queries will leverage the index on \texttt{Tenant_ID}.
For the second workload, it marks the first index as redundant since these
queries will leverage the second and third indexes.
Thus, \detector supports complex rules.
\end{example}
\vspace{-1em}

\section{Ranking Anti-Patterns}
{\label{sec:ranking}}

In this section, we present the algorithm used by \ranker for ordering the
\apsshort identified by \detector.
We begin with an overview of the metrics collected by \ranker for ordering the
\apsshort.
We then present the model used by \ranker.

\begin{figure}[t]
\centering
\subfloat[Role Table]{
\footnotesize
\begin{tabular}{r c }
                   \textbf{Role\_ID}  & \textbf{Role\_Name} \\
    \hline
    1 & R1 \\
    2 & R2 \\
    3 & R3 \\ 
\end{tabular}

\label{fig:role-table}
}
\hfill
\subfloat[User Table]{
\footnotesize
\begin{tabular}{r c c c }
                   \textbf{User\_ID}  & \textbf{Name} &
                   \textbf{Role} & \textbf{Email}
    \\
	\midrule
    U1  & N1  & 1 &  E1 \\
    U2  & N2  & 2 &  E2 \\
    U3  & N3  & 2 &  E3 \\        
    U4  & N4  & 3 &  E4 \\        
\end{tabular}

\label{fig:user-with-role}
}

\caption{
\textbf{Refactored GlobaLeaks Application} --
List of tables.
}
\label{fig:user-role-relation}
\end{figure}


\begin{figure}
\centering

\begin{tabular}{c}
$\mathcal{S}_{rp}(x)$, $\mathcal{S}_{wp}(x)$, $\mathcal{S}_{m}(x)$  = $\min{(1, x/5)}$\\
$\mathcal{S}_{da}(x)$  = $min(1, x/8)$ \\
$\mathcal{S}_{di}(x)$, $\mathcal{S}_{a}(x)$  = $x$  // $x$ $\in$ $\{0, 1\}$ \\
\\

\texttt{score} = 
$\mathcal{W}_{rp}$ * $\mathcal{S}_{rp}$($\mathcal{RP}$) + 
$\mathcal{W}_{wp}$ * $\mathcal{S}_{wp}$($\mathcal{WP}$) + \\
$\mathcal{W}_{m}$ * $\mathcal{S}_{m}$($\mathcal{M}$)  + 
$\mathcal{W}_{da}$ * $\mathcal{S}_{da}$($\mathcal{DA}$) + \\
$\mathcal{W}_{di}$ * $\mathcal{S}_{di}$($\mathcal{DI}$) + 
$\mathcal{W}_{A}$ * $\mathcal{S}_{a}$($\mathcal{A}$) \\

\end{tabular}


\caption{\textbf{Ranking Model --} Formulae for measuring the impact of
\apsshort.}
\label{fig:ranking-formula}
\end{figure}

\begin{figure}
\small
\subfloat[Ranking model configurations]{
\centering
\begin{tabular}{r c c c c c c c}
                   & $\mathcal{W}_{rp}$  & $\mathcal{W}_{wp}$ & $\mathcal{W}_{m}$ &
                   $\mathcal{W}_{da}$ & $\mathcal{W}_{di}$ & $\mathcal{W}_{a}$ &                   
    \\
	\midrule

    \textit{C1}  & 0.7 & 0.15 & 0.05 & 0.04 & 0.02 & 0.02  \\
    \textit{C2}  & 0.4 & 0.4 & 0.1 & 0.04 & 0.02 & 0.02  \\
\end{tabular}

\label{fig:ranking-config}
}
\hfill
\subfloat[Impact of \apsshort ]{
\centering
\begin{tabular}{l c c c c c c}
   & $\mathcal{S}_{rp}$  & $\mathcal{S}_{wp}$  
   & $\mathcal{S}_{m}$ & $\mathcal{S}_{da}$ 
   & $\mathcal{S}_{di}$ & $\mathcal{S}_{a}$ 
    \\
	\midrule

    Index Underuse  & 1.5x & 0 & 0 & 0 & 0 & 0 \\
    Enumerated Types  & 0 & >10x & 2 & 1 & 0 & 0 \\
\end{tabular}

\label{fig:ranking-ap-impact}
}

\caption{\textbf{Ranking Model Configurations --} Illustration of the impact of
the ranking model configuration on the ordering of \apsshort.}
\label{fig:ranking-example}
\end{figure}
\subsection{Metrics for Ranking Anti-Patterns}
\label{sec:ranking::metrics}

\ranker collects six metrics for each \apshort. 
These metrics are subsequently used by the model for ordering the
\apsshort.

\dcircle{1} 
\textbf{Read and Write Performance (RP, WP):}
This metric characterizes the impact of the \apshort on the application's
performance.
We measure the time taken to execute different types of frequently-observed
queries in the presence and absence of each \apshort.
For this analysis, we focus on the following types of queries:
(1) a lookup query that retrieves a set of records from a table based on a
highly selective predicate (\texttt{SELECT}),
(2) an aggregation query that computes the sum of all the elements in a column
(\texttt{SUM}),
(3) a join query that combines the records in two tables in an application based
on a join predicate (\texttt{JOIN})), and
(4) an update statement that modifies a set of records in a table based on a
highly selective predicate (\texttt{UPDATE}). 
\ranker uses the results of this quantitative analysis to estimate the potential
speedup in executing the target application's queries by fixing an \apshort.
For example, fixing the multi-valued attribute \apshort
(\cref{ex:multi-valued-attribute}) accelerates lookup and join queries by
636$\times$ and 256$\times$, respectively.

\dcircle{2} 
\textbf{Maintainability (M):}
The next metric appraises the impact of each \apshort on the maintainability
of the application.
We conduct a qualitative analysis of the \textit{number of changes}
($\mathcal{C}$) needed in an application to support a new task in the presence
and absence of each \apshort.
This determines the degree of refactoring necessitated by the \apshort.
If $\mathcal{C}$ is linearly or super-linearly dependent on the \textit{number
of queries} ($\mathcal{Q}$) present in the application, then \ranker will
prioritize this \apshort.
In contrast, a design wherein $\mathcal{C}$ is independent of $\mathcal{Q}$ 
improves the extensibility of the application.

Consider the enumerated types \apshort (\cref{ex:enumerated-types}).
If the developer would like to rename a particular \texttt{Role} 
(\eg, \texttt{R2} $\mapsto$ \texttt{R5}), they would need to
execute the following queries:

\begin{lstlisting}[style=SQLStyle]
ALTER TABLE User DROP CONSTRAINT IF EXISTS User_Role_Check;
UPDATE User SET Role='R5' WHERE Role='R2';
\end{lstlisting}

\cref{fig:user-with-role} illustrates an alternate design wherein only 
one query (\ie, $\mathcal{C}$ = $\mathcal{O}$($1$)) is sufficient for this
refactoring.

\begin{lstlisting}[style=SQLStyle]
UPDATE Role SET Role_Name='R5' WHERE Role_Name='R2';
\end{lstlisting}

\dcircle{3} 
\textbf{Data Amplification (DA):}
The next metric appraises the impact of \apsshort on \textit{data
amplification}.
\apshort-free design can shrink the storage footprint of an application.

Consider the enumerated types \apshort in \gl (\cref{ex:enumerated-types}).
The \texttt{Role} column can take the following \texttt{STRING} values: 
(\texttt{R1}, \texttt{R2}, and \texttt{R3}).
Repeatedly storing these \texttt{STRING} values increases the storage footprint
of the application.
The alternate design presented in~\cref{fig:user-with-role} addresses this
limitation by introducing a \texttt{ROLE} table and encoding the \texttt{Role}
column in the \texttt{USER} table using \texttt{INTEGER} values: (1, 2, and 3).
In addition to reducing data amplification, it allows the developer to
utilize foreign key constraints (\eg, to ensure that every user can only take
on one of these roles)
\footnote{This relationship can also be preserved using a \texttt{CHECK}
constraint~\cite{pg-check}. However, this feature reduces performance and 
maintainability (\autoref{evaluation::ranking-repair}).}.

\dcircle{4} 
\textbf{Data Integrity (DI):}
The third metric characterizes the impact of each AP on data integrity.
We examine how an \apshort affects the integrity of the application. 
For instance, consider the multi-valued attribute \apshort in \gl
(\cref{ex:multi-valued-attribute}).
If \new{a}   user with \texttt{User\_ID} \texttt{u1} is deleted from the \texttt{USER}
table, we need to manually execute another query to delete the \texttt{u1}
string from the comma-separated \texttt{User\_IDs} field.

\begin{lstlisting}[style=SQLStyle]
UPDATE Tenants SET User_IDs = REPLACE(User_IDs, ',u1', '') 
WHERE User_IDs LIKE '%u1%';
\end{lstlisting}

If this query is not executed, the data integrity constraint will be violated.
In contrast, if the database contains an intersection table as shown
in~\cref{fig:pattern-1-good-3}, we can leverage DBMS features for preserving
integrity constraints (\eg, cascaded deletes as shown
in~\autoref{sec:casestudy}).

\dcircle{5}
\textbf{Accuracy (A):} 
This metric characterizes the impact of the \apshort on the \textit{accuracy} of
the returned results.
Consider the no foreign key \apshort (\cref{ex:no-foreign-key}). 
In the \texttt{QUESTIONNAIRE} table, since there is no foreign key linking the
\texttt{Tenant\_ID} columns in the \texttt{Tenant} and \texttt{QUESTIONNAIRE}
tables, delete operations will not be cascaded.
The resultant dangling references lead to tuples with \texttt{NULL} values 
when  these tables are joined.


\subsection{Model for Ranking Anti-Patterns}
\label{sec:ranking::model}

We now present the ranking model that \ranker uses for ordering the \apsshort
identified by \detector.
Our goal is to prioritize the attention of developers on high-impact \apsshort.
The model leverages the metrics presented in~\autoref{sec:ranking::metrics}.

\ranker sorts the \apsshort in the application in decreasing order of their
estimated impact on performance, maintainability, and accuracy.
To do this estimation, it maps the queries in the application to the standard
types of queries that have already been evaluated.
It then generates a query-aware ranking of \apsshort in the application.
The developer can tailor the weights used by the model for these different
features: performance, maintainability, and accuracy.
It then sends the ordered list of \apsshort to \fixer.
For \apsshort with multiple candidate fixes, \fixer suggests the best 
fix based on the collection of queries present in the application.
We defer a discussion of \fixer to~\autoref{sec:repair}.
As new performance data is collected over time, we update the ranking model 
to improve the quality of its decisions.

\PP{Model Components:}
The model consists of two components: 
(1) \textit{intra-query} and (2) \textit{inter-query} ranking components. 
The intra-query component ranks the \apsshort detected in each query.
It first computes the following metrics for each \apshort:
(1) read performance ($\mathcal{RP}$), 
(2) write performance ($\mathcal{WP}$), 
(3) maintainability ($\mathcal{M}$), 
(4) data amplification ($\mathcal{DA}$),
(5) data integrity ($\mathcal{DI}$), and
(6) accuracy ($\mathcal{A}$).

It then aggregates these metrics using the weights
shown in~\cref{fig:ranking-formula}.
The developer can configure these weights to best meet their applications
requirements.
For instance, if an application requires higher read performance, 
the developer can increase the \textit{read performance} weight.
\ranker uses the computed aggregate score for ranking the \apsshort within a
query and to compute the score for each \apshort.

The inter-query component sorts \apsshort based on their impact on all the
queries in the application.
The developer can choose one of two inter-query ranking models: 
\dcircle{1} based on number of \apsshort in each query (\ie, queries with 
more \apsshort are ranked higher), or 
\dcircle{2} based on the computed score.

\begin{example}
\label{ex:ranking}
Consider a query suffering from the \textit{index underuse} and
\textit{enumerated types} \apsshort.
\cref{fig:ranking-config} illustrates two different configurations of the
ranking model (\textit{C1} and \textit{C2}).
\cref{fig:ranking-ap-impact} lists the metrics associated with the detected
\apsshort.
The first configuration (\textit{C1}) prioritises read performance
((\eg, analytical workloads).
So, it ranks the index underuse \apshort (\texttt{score} = $0.21$) higher
than the enumerated types \apshort (\texttt{score} = $0.175$).
In contrast, the second configuration (\textit{C2}) gives equal priority to both
read and write performance (\eg, hybrid transactional/analytical workloads).
So, it ranks the enumerated types \apshort (\texttt{score} = $0.47$) higher 
than the index underuse types \apshort (\texttt{score} = $0.12$).
In this manner, \ranker allows the developer to prioritise \apsshort.
\end{example}


\PP{Conflicting Fixes:}
\new{
Fixes for \apsshort detected in an application may conflict with each other.
\ranker assists the developer in resolving these conflicts by prioritising the 
\apsshort.
For instance, consider an application with these two \apsshort: \tmj and \ent.
To resolve the latter \apshort, the developer must create a new table for the
attribute with an enumerated type (\eg, \texttt{ROLE} table).
However, this fix would increase the performance impact of former \apshort as it
would require the developer to connect the newly added table with an additional
\texttt{JOIN} in \texttt{SELECT} queries.
\sys orders the detected \apsshort based on the user-specified ranking model.
So, if the developer is prioritising read performance, then they may fix the
former \apshort first and ignore the latter one.
In this manner, they can iteratively fix the \apsshort in the application based
on their impact score from \ranker.
}

\section{Fixing Anti-Patterns}
{\label{sec:repair}}

Merely identifying the high-impact \apsshort will not be sufficient since
application developers who are experts in other domains are likely not 
familiar with anti-patterns~\cite{data-science,segaran09}.
\fixer addresses this problem by automatically suggesting alternate database
designs and queries that are tailored to the application.
We begin with an overview of the algorithm used by \fixer.
We then describe the query repair engine that \fixer leverages for rewriting
\sql queries in~\autoref{sec:repair-engine}.

\PP{Overview:}
As shown in~\cref{algo:fixing-ap}, \fixer takes the following inputs:
(1) a list of detected \apsshort,
(2) the parse trees of the queries containing those \apsshort, and 
(3) the context of the application.
Depending on the types of \apsshort, it fetches the associated rules for fixing
them.
Besides targeting the queries containing the \apsshort,
\fixer retrieves the list of queries $\mathcal{I}$  that are also impacted by
the \apshort fix from the application's context.
It appends these impacted queries to the list of queries containing \apsshort to
construct the list of queries that must be transformed ($\mathcal{Z}$).
\fixer then passes this list to the \textit{query repair} engine
(\cref{algo-line:apply-rule}).
The rule engine checks whether it can generate non-ambiguous query 
transformations for a given query based on the \apsshort that it contains.
If that is the case, then it applies those transformations on the query's parse
tree (\cref{algo-line:query-transform}).
It then transforms the parse tree to a SQL string based on the dialect used by 
the application.
If it cannot generate non-ambiguous transformations, then it returns a 
textual fix that is tailored based on the context
(\cref{algo-line:text-response}).
The application developer must subsequently follow the guidance provided in the
textual fix to manually resolve the detected \apsshort.

\begin{algorithm}[t]
\SetKwFunction{rulesFor}{\textsc{Get}\textsc{Rules}\textsc{For}\textsc{Anti}\textsc{Pattern}}
\SetKwFunction{ruleEngine}{\textsc{Rule}\textsc{Engine}}
\SetKwFunction{getTransformations}{\textsc{Get}\textsc{Transformations}}

\SetKwFunction{statsFor}{\textsc{Stats}}
\SetKwFunction{resolveAP}{\textsc{Resolve}}
\SetKwFunction{detectionRule}{\textsc{Rule}}
\SetKwFunction{transform}{\textsc{Transform}}
\SetKwFunction{textResponse}{\textsc{Get}\textsc{Textual}Fix}
\SetKwFunction{toSQL}{To\textsc{sql}}
\SetKwFunction{impacted}{\textsc{Get}\textsc{Impacted}Queries}
\SetAlgoLined
\SetKwInOut{Input}{input} 
\SetKwInOut{Output}{output}
\caption{Fixing Anti-Patterns}
\label{algo:fixing-ap}

\Input{detected anti-patterns $\mathcal{P}$, 
parsed queries $\mathcal{Q}$,
context $\mathcal{C}$}
\Output{anti-pattern fixes $\mathcal{F}$}

    fixes $\mathcal{F}$ $\gets$ $\{$ $\}$ \\
	\For{anti-pattern $p$ \textbf{in} $\mathcal{P}$}{
		fix rules $\mathcal{R}$ $\gets$ \rulesFor{$p$}

	    \tcp{Identify queries which are impacted by the anti-pattern fix} 
	    impacted-queries $I$ $\gets$ \impacted{$p$, $\mathcal{C}$}
	    \label{algo-line:impacted-queries} \\
	    
	    to-be-transformed-queries $Z$ $\gets$ $\mathcal{Q}$ $\cup$ $I$

	    \tcp{Pass to-be-transformed queries to the query repair engine} 	    
	    \For{query $z$ \textbf{in} $\mathcal{Z}$}{
	    
	    	query transformations $\mathcal{T}$ $\gets$ \getTransformations($z$, $p$) 
	    	\label{algo-line:apply-rule} \\
	    
		    \uIf{$\mathcal{T}$ is not empty}{
		        transformed-parsed-query $t$ $\gets$ \transform{$z$, $\mathcal{T}$}
		        \label{algo-line:query-transform}\\
		        fixed-sql-query $f$ $\gets$ \toSQL{$t$}
		    } \Else{
		        \tcp{Return a textual fix tailored for application}
		        textual fix $f$ $\gets$ \textResponse{$p$, $z$}
		        \label{algo-line:text-response} 
		    }
		    
		    $\mathcal{F}$.append($f$)		    	    	    
	    }	    	
	}
 \textbf{return} $\mathcal{F}$ 
\end{algorithm}

\subsection{Query Repair Engine}
\label{sec:repair-engine}

The query repair engine transforms a given \sql statement based on a set of
\textit{rules} for fixing \apsshort.
The rule system is instrumental in facilitating our experimentation with
statement transformations for two reasons. 
First, the rule system paradigm makes it easy for \fixer to exploit the
complicated triggering interactions between the repair rules, 
thereby obviating the need for explicitly laying out the flow of control 
between rules. 
Second, the rule system is extensible.
This extensibility allowed us to formulate and evaluate tens of transformations
over time.

In addition to rewriting existing \sql statements, \fixer also needs to
construct new statements for certain \apsshort. 
For example, in the case of the multi-valued attribute \apshort
(\autoref{ex:multi-valued-attribute}), \fixer first 
constructs a new \texttt{HOSTING} table and then updates the schema of 
the \texttt{TENANTS} table, as shown below:

\begin{lstlisting}[style=SQLStyle]  
/* Create an intersection table */
CREATE TABLE Hosting ( 
    User_ID VARCHAR(10) REFERENCES Users(User_ID), 
    Tenant_ID VARCHAR(10) REFERENCES Tenants(Tenant_ID)
);
/* Drop redundant column */
ALTER TABLE Tenants DROP COLUMN User_IDs;
\end{lstlisting}

A key challenge for the query repair engine is that it must identify all the
queries which are impacted by the anti-pattern fix and transform them as well.
For instance, with the intersection table, \fixer rewrites 
the query for retrieving information about the users served by tenant thus:

\begin{lstlisting}[style=SQLStyle] 
/* Retrieve information about users served by tenant */
SELECT * FROM Hosting as H JOIN Tenants as T
ON H.User_ID == T.User_ID WHERE H.Tenant_ID = 'T1';
\end{lstlisting}

\PP{Rule Representation:}
We represent rules in our engine as pairs of functions in a procedural language.
Each rule consists of: 
(1) a \textit{detection function} (\autoref{sec:detection}), 
which does an arbitrary check and sets a flag \texttt{TRUE} or \texttt{FALSE},
and 
(2) an \textit{action function}, which, if the condition function sets the 
flag \texttt{TRUE}, is invoked to take an arbitrary action, 
such as transforming existing \sql statements and creating new \sql statements.



\section{Implementation}
{\label{sec:implementation}}

\sys is implemented in Python ~\cite{python} and exports the following three interfaces.
%
(1) Interactive Shell, (2) REST, and (3) GUI.
Application developers and \sql IDE developers can leverage these interfaces to either directly interact with \sys or to integrate it with their own IDEs.
We describe these interfaces below:
%

\squishitemize

\item \textbf{Interactive Shell:}
An \sql application developer can import the \sys package from a package
repository (\eg, PyPI ~\cite{pypi}) and directly use the interactive shell
interface to execute \sql queries or leverage these sub-modules in other
tools.

\begin{lstlisting}[style=PythonStyle]
# Import the anti-pattern finder method
from sqlcheck.finder import find_anti_patterns
query = `INSERT INTO Users VALUES (1, 'foo')`
results = find_anti_patterns(query)
\end{lstlisting}

\item \textbf{REST Interface:} 
This interface allows developers to leverage \sys in applications 
developed in other programming languages by using web requests via HTTP.
We implement this using the Flask web framework~\cite{python-flask}.

\begin{lstlisting}[style=JSONStyle]
HTTP POST /api/check
Body: {"query":"INSERT INTO Users VALUES (1,'foo')"}
\end{lstlisting}


 \item \textbf{GUI Interface:} 
Lastly, this interface is geared towards a wider range of users who are not familiar with application programming.
This interface enables users to easily get feedback on their queries by
copying them into the \texttt{input} field and is developed using
ReactJS~\cite{reactjs}.
%
%
%
\squishend 

\PP{Extensibility:} 
\sys is extensible by design.
A developer may add a new \apshort rule that implements the generic rule
interface (name, type, detection rule, ranking metrics, and repair rule) 
and register it in the \sys rule registry.
A developer may also extend the context builder to augment the application's
context for supporting complex rules.
Lastly, a developer may replace the non-validating parser with a DBMS-specific
parser to increase the utility of the parse tree.

\section{Evaluation}
{\label{sec:evaluation}}

We evaluated \sys on a variety of real-world \sql queries 
to quantify its efficacy in detecting, ranking and fixing \apsshort.
We illustrate that:

\squishitemize
  \item \textbf{Detection:} 
  \sys detects a wider range of \apsshort (\new{26}) in real-world DBMS applications 
  compared to \dbdeo (11).
  \detector has \texttt{48\%} fewer false positives and  \texttt{20\%} fewer
  false negatives than \dbdeo resulting in higher precision and recall.
  \new{
  \sys found \djangomajorapcount major \apsshort in \djangoappcount real-world 
  web applications.
  }
   
  \item \textbf{Ranking:} 
  \ranker allows the developer to order \apsshort based on their impact.
  Its ranking model is derived through an empirical analysis of 
  \globaleaks
.
  We show that the average and maximal impact of \apsshort on 
  runtime
  performance is 4$\times$ and  $>$10000$\times$, respectively.

  \item \textbf{Fixing:}
  We conduct a user study to validate the efficacy of \sys in fixing \apsshort.
  \sys's efficacy in resolving the \apsshort in the queries written by users is
  51\%. 
  Overall, the participants of the study confirmed that \sys helped eliminate
  \apsshort in their applications.

\squishend

\subsection{Detection of Anti-Patterns} 
\label{evaluation::detection}

\begin{table}[t]
  \centering
  \footnotesize
  \begin{tabular}{r | c c c | c c c c}
    \textbf{\apshort Name}  &  \textbf{S} & \textbf{D}  & \textbf{Both} &
    \textbf{TP-S} & \textbf{FP-S} & \textbf{TP-D} & \textbf{FP-D}
\\
\hline
Pattern Matching & 1037 & 524 & 28 & 705 & 332 & 0 & 524 \\
God Table & 27 & 170 & 3344 & 27 & 0 & 0 & 170 \\ 
Enumerated Types & 414 & 42 & 48 & 411 & 3 & 0 & 43 \\ 
Rounding Errors & 352 & 7 & 1074 & 329 & 23 & 0 & 7 \\
Data in Metadata & 18 & 584 & 1226 & 18 & 0 & 93 & 491 \\
Adjacency List & 0 & 10 & 93 & 0 & 0 & 0 & 10 \\
\hline
\textbf{Total:}& 1848 & 1337 & 5813 & 1588 & 358 & 93 & 3783  
\end{tabular}

  \caption{\textbf{Detection of Anti-Patterns:} 
  Comparison of the number of \apsshort identified by \sys and \dbdeo in the
  query benchmark.
  Columns \textbf{S} and \textbf{D} report the \apsshort detected by
  \textit{only} that tool.
  Column \textbf{Both} lists the \apsshort detected by both tools.
  \textbf{TP} and \textbf{FP} refer to true and false positives, respectively.
  }  
  \label{tab:sqlcheck-vs-dbdeo}
\end{table}

To our knowledge, \dbdeo is the closest system to \sys.
\dbdeo is effective in uncovering \apsshort in real-world \sql queries.
However, \dbdeo's query analysis algorithm suffers from low precision and
recall.
Furthermore, it differs from \sys in that it neither ranks the 
\apsshort based on their impact nor suggest fixes for the detected \apsshort.
In this experiment, we compare the \apshort-coverage and
accuracy of \sys against that of \dbdeo.

\PP{Query Benchmark:} 
We download \appcount open-source repositories containing \sql 
statements from GitHub~\cite{github}.
\new{
We extract around \texttt{174} thousand string-quoted embedded 
\sql statements from these repositories.
We then use regular expressions to extract \sql statements from the files
contained in these repositories.
%
%
%
%
%
}
We evaluate \sys under two different configurations: (1) with only intra-query
analysis, and (2) with both intra- and inter-query analyses.
The open-source applications hosted on GitHub only contain \sql queries 
and not their associated databases.
So, \sys cannot leverage its \textit{data analysis} techniques in this
experiment.

\PP{Results:} 
We group and aggregate the detected \apsshort based on their type. 
The results (details in ~\cref{tab:ap-distribution-github-survey-dbdeo-sqlcheck}) are as follows:

\squishitemize
\item \dbdeo detects \texttt{14764} \apsshort (\texttt{11} types of \apsshort).
\item \sys (only intra-query analysis) detects \texttt{86656} \apsshort (\texttt{18}
types of \apsshort). 
\item \sys (intra- and inter-query analysis) detects \texttt{63058} \apsshort (\texttt{\new{21}} types of \apsshort). 
\squishend

\squishitemize
  \item \textbf{Coverage and Accuracy:}
  Under both configurations, \sys detects a wider range of \apsshort 
  compared to \dbdeo.
  With only intra-query analysis enabled, \sys finds 2.6$\times$ more
  \apsshort than \dbdeo.
  We attribute this increase in recall to two factors.
  First, \sys supports \totalapsupportcount types of \apsshort (\dbdeo only supports 11
  types.)
  \new{
  Second, \sys uses detection rules that are capable of uncovering
  different variants of the same \apshort (\eg, set of regular expressions for
  identifying \mva:}
  \verb|(id\\s+regexp)||\verb|(id\\s+like)|).
  This increase in recall also results in higher false positives 
  (\ie, lower precision).
  Enabling both intra- and inter-query analyses mitigates this problem. 
  Under this configuration, \sys reports three additional types of \apshort but 1.8 $\times$ fewer \apsshort
  compared to the prior configuration.
  This is because it eliminates false positives by leveraging the inter-query context.
   
  \item \textbf{Dialect-Coverage:}
  Qualitatively, both \sys and \dbdeo support a wide range of \sql dialects.
  We attribute this to their usage of \parser, a non-validating parser,
  that supports diverse dialects.
  Furthermore, \sys leverages \texttt{sqlalchemy} to construct the query and
  context objects in a DBMS-agnostic manner~\cite{sqlalchemy}.
\squishend

We next conduct a manual analysis of the \apsshort reported by \dbdeo and \sys
in the query benchmark for a subset of \apsshort.
We do not examine certain \apshort (\eg, \npk, \ixu) because \dbdeo does not
report the query in which the \apshort was detected.
The results are shown in~\cref{tab:sqlcheck-vs-dbdeo}.
\sys has \texttt{48\%} fewer false positives and \texttt{20\%} fewer false negatives compared
to \dbdeo.
This illustrates the impact of intra- and inter-query analyses in increasing
the precision and recall of \apshort-detection.

%
\subsection{Ranking and Repair of Antipatterns}
\label{evaluation::ranking-repair}

We next examine the impact of \apsshort on runtime performance.
In particular, we compare the query execution time before and after fixing a
given \apshort in a real-world application.

\PP{Experiment Setup:}
We aggregate the \apsshort detected in the query benchmark presented
in~\autoref{evaluation::detection} based on their associated application.
We rank these applications based on the frequency and types of detected
\apsshort.
Based on this ranking, we select the \globaleaks application for this
experiment~\cite{globaleaks}.

\globaleaks leverages \texttt{sqlalchemy} (an object-relational mapping (ORM)
framework)~\cite{sqlalchemy}.
We first transform the ORM operations to \sql queries.
We then recreate the database schema on a DBMS instance (PostgreSQL v11.2) and
load a synthetic dataset containing 10~M records (19~GB across 11 tables).
\globaleaks inherently contains ten types of \apsshort.
We infuse three additional \apsshort for quantifying their performance
impact.
For instance, we add comma-separated strings in a column to infuse the
multi-valued attribute \apshort.

We quantify the performance impact of every \apshort.
For each \apshort, we execute different types of queries in \globaleaks under
two configurations: (1) before the \apshort is fixed, and (2) after the
\apshort is fixed using the feedback provided by \fixer.
For each query, we report the average execution time of five runs.

\begin{figure*}[t]
\captionsetup[subfigure]{justification=centering}
\centering
\subfloat[][Index Overuse:\\ Update]{
\centering
\includegraphics[width=0.12\textwidth]{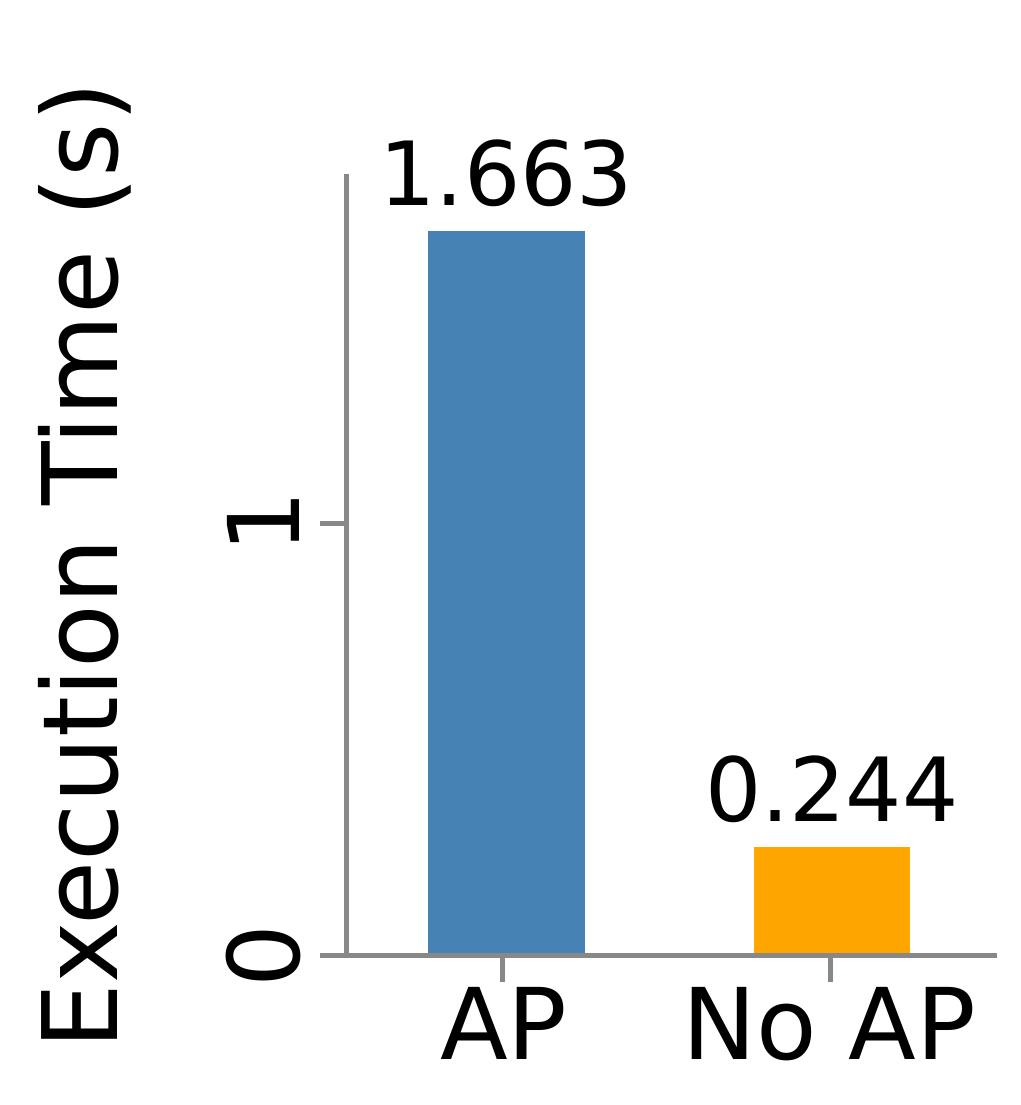}
\label{fig:index-over-use-update}
}
\qquad
\subfloat[][Index Underuse:\\ Grouped Aggregate]{
\includegraphics[width=0.12\textwidth]{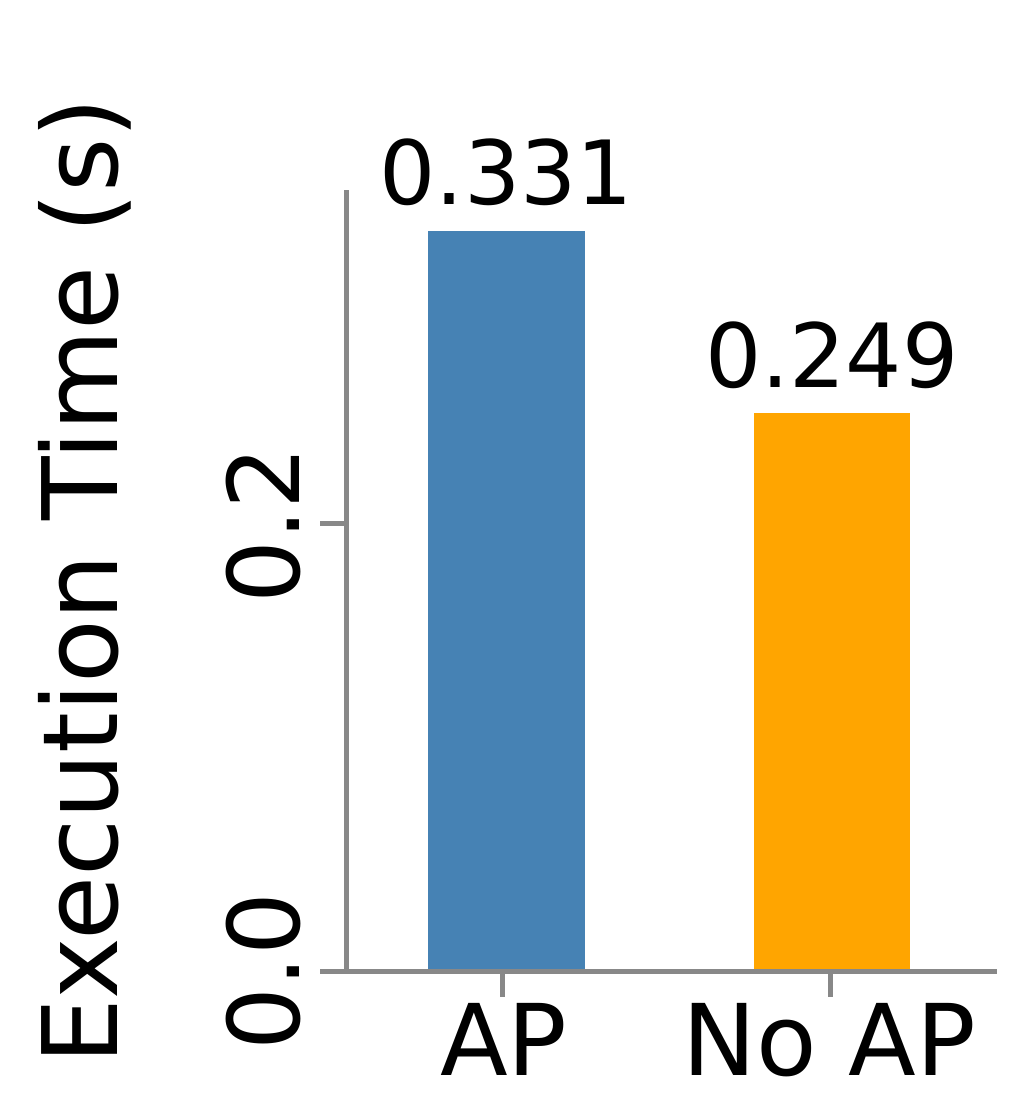}
\label{fig:index-underuse-select-aggr-wo-predicate}
}
\qquad
\subfloat[][Index Underuse:\\ Scan with Predicate]{
\centering
\includegraphics[width=0.12\textwidth]{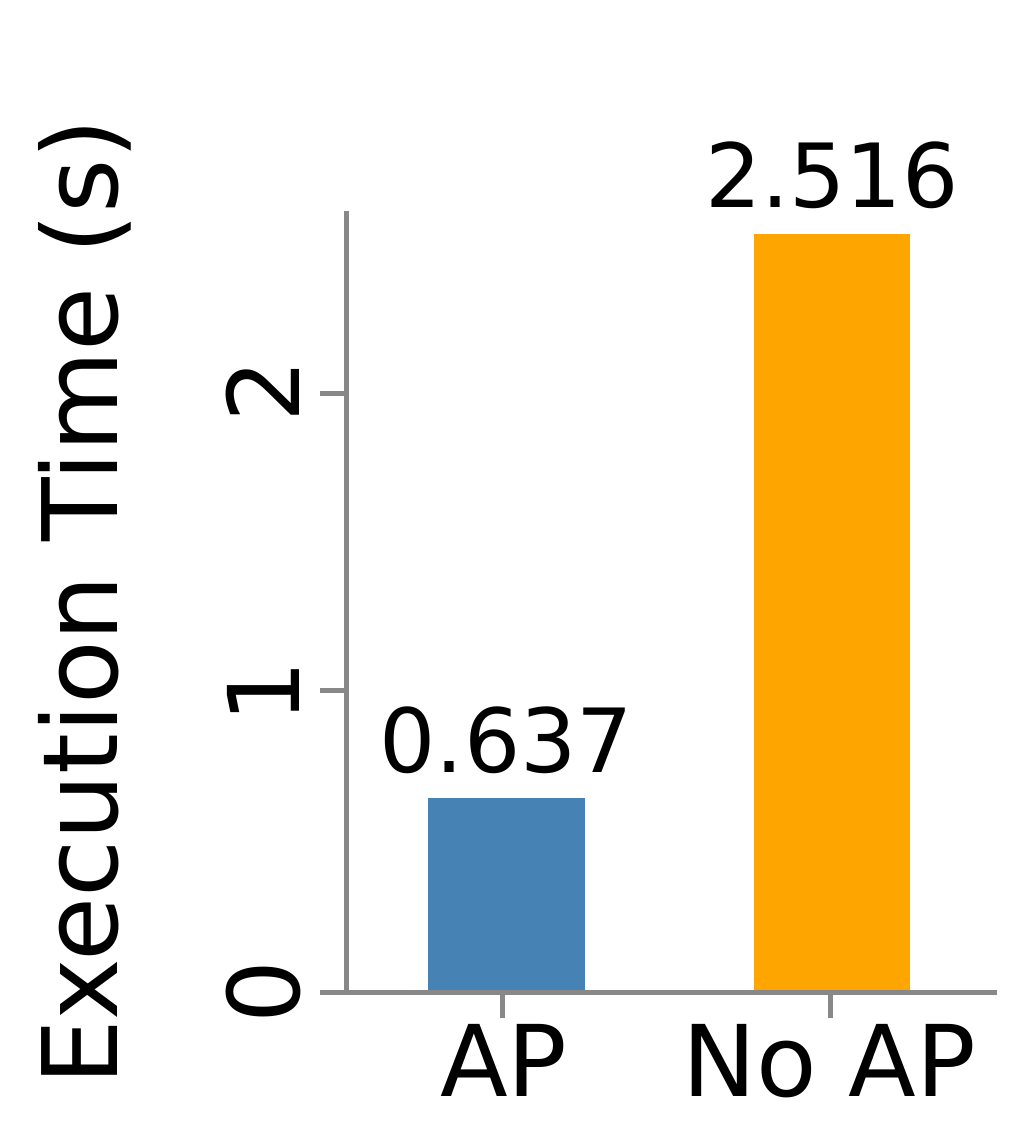}
\label{fig:index-underuse-select-predicate}
}
\qquad
\subfloat[][Foreign Key Exists:\\ Update]{
\centering
\includegraphics[width=0.12\textwidth]{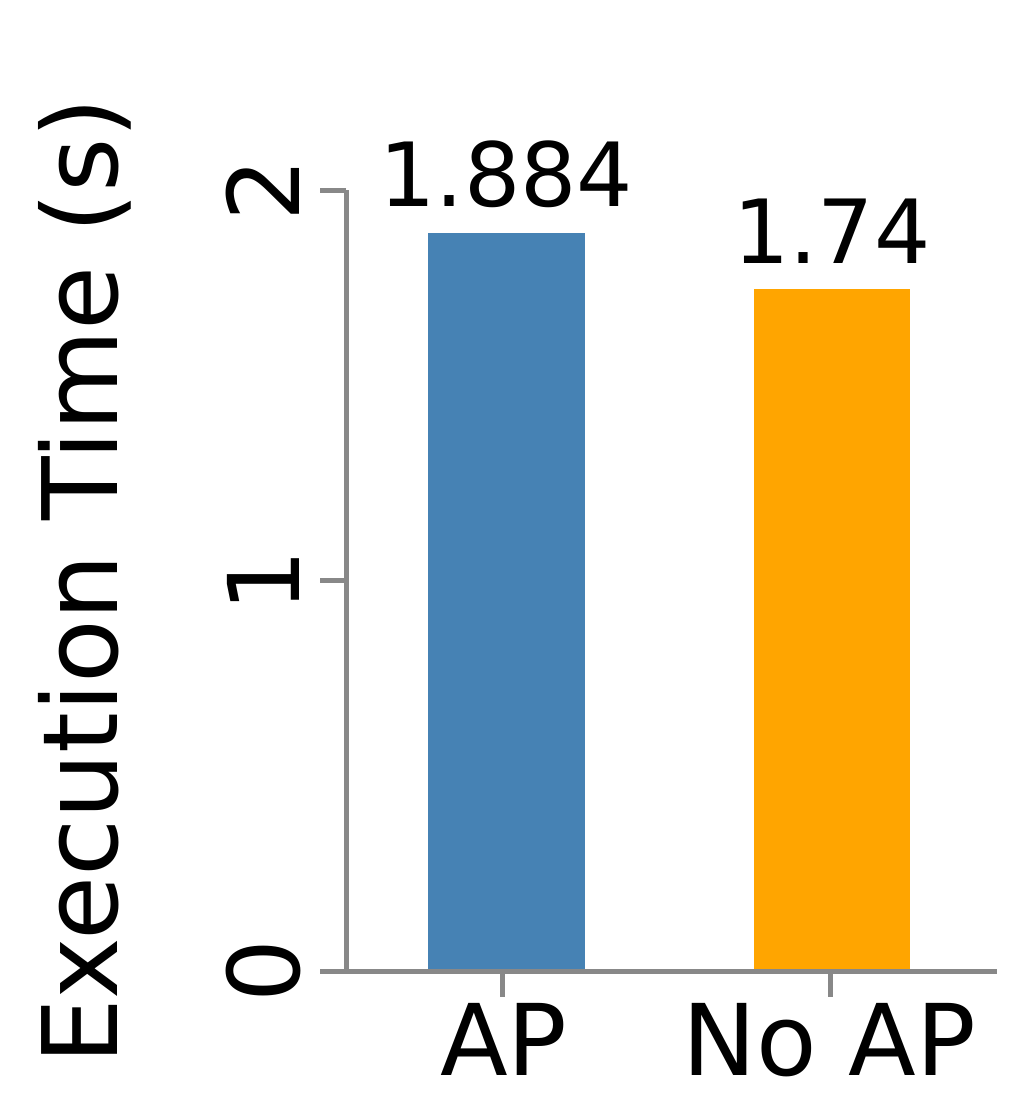}
\label{fig:fk-exists-update}
}
\qquad
\subfloat[][Foreign Key Exists:\\ Select]{
\centering
\includegraphics[width=0.12\textwidth]{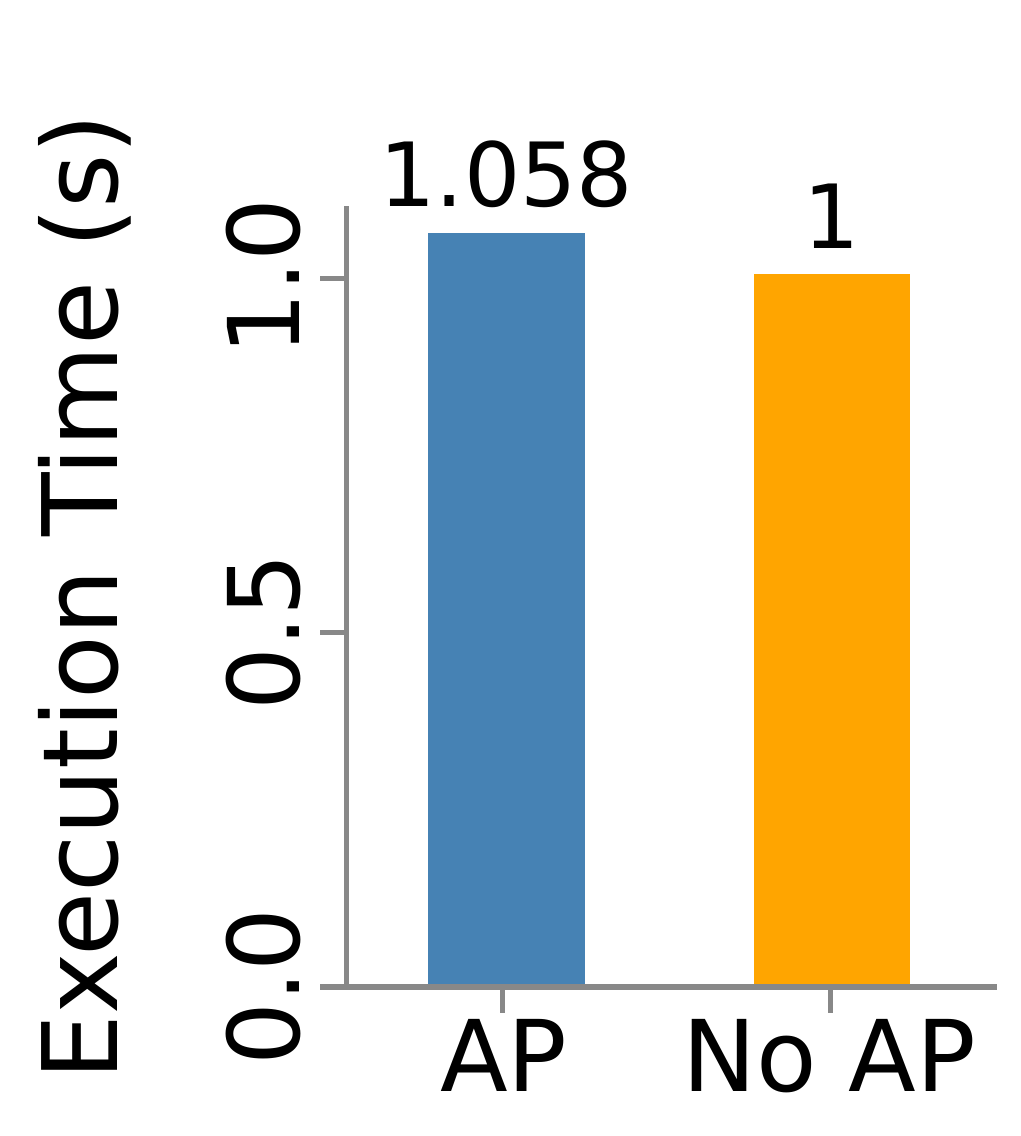}
\label{fig:fk-exists-select}
}
\qquad
\subfloat[][Foreign Key Exists:\\ Update with Index]{
\centering
\includegraphics[width=0.12\textwidth]{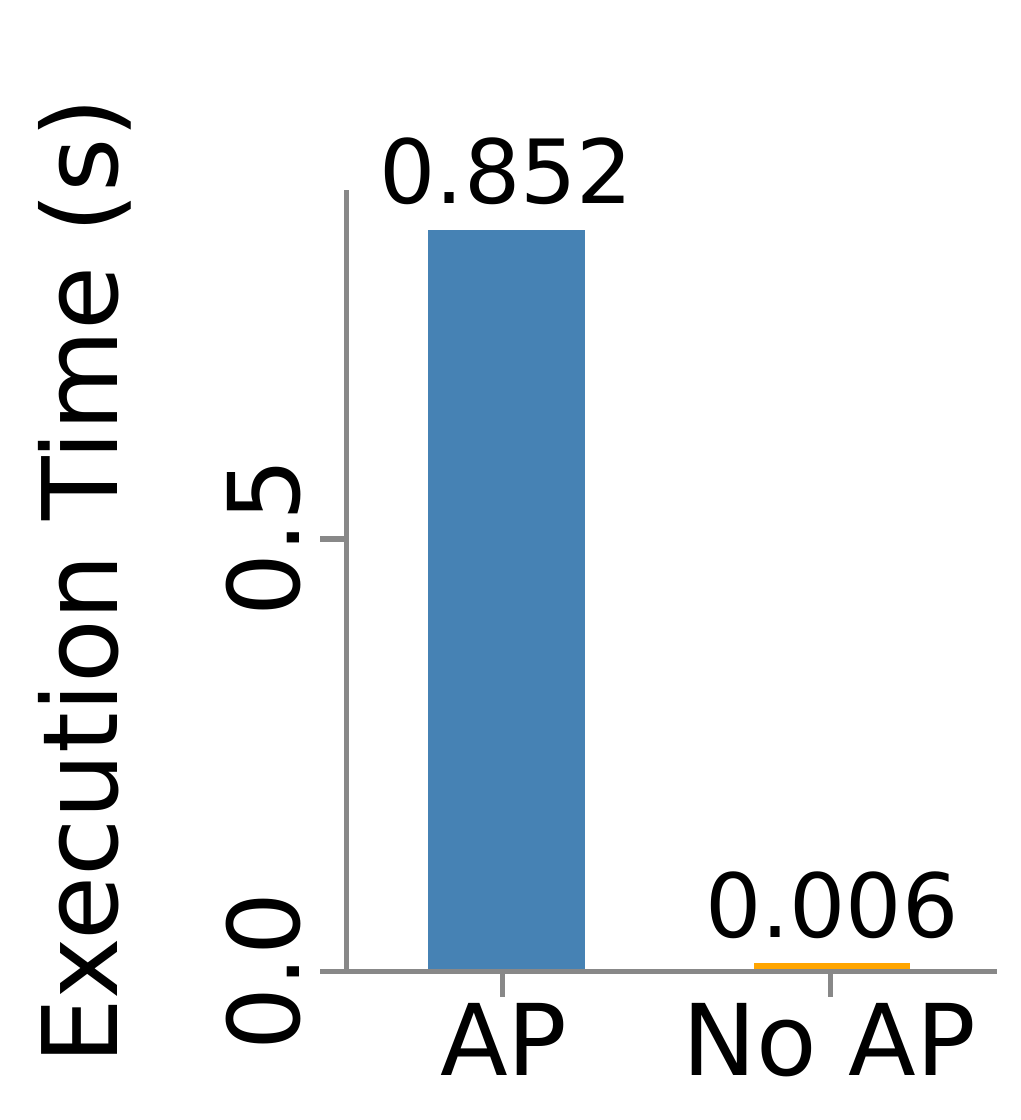}
\label{fig:fk-exists-update-index}
}
\qquad
\subfloat[][Enumerated Types:\\ Update]{
\centering
\includegraphics[width=0.12\textwidth]{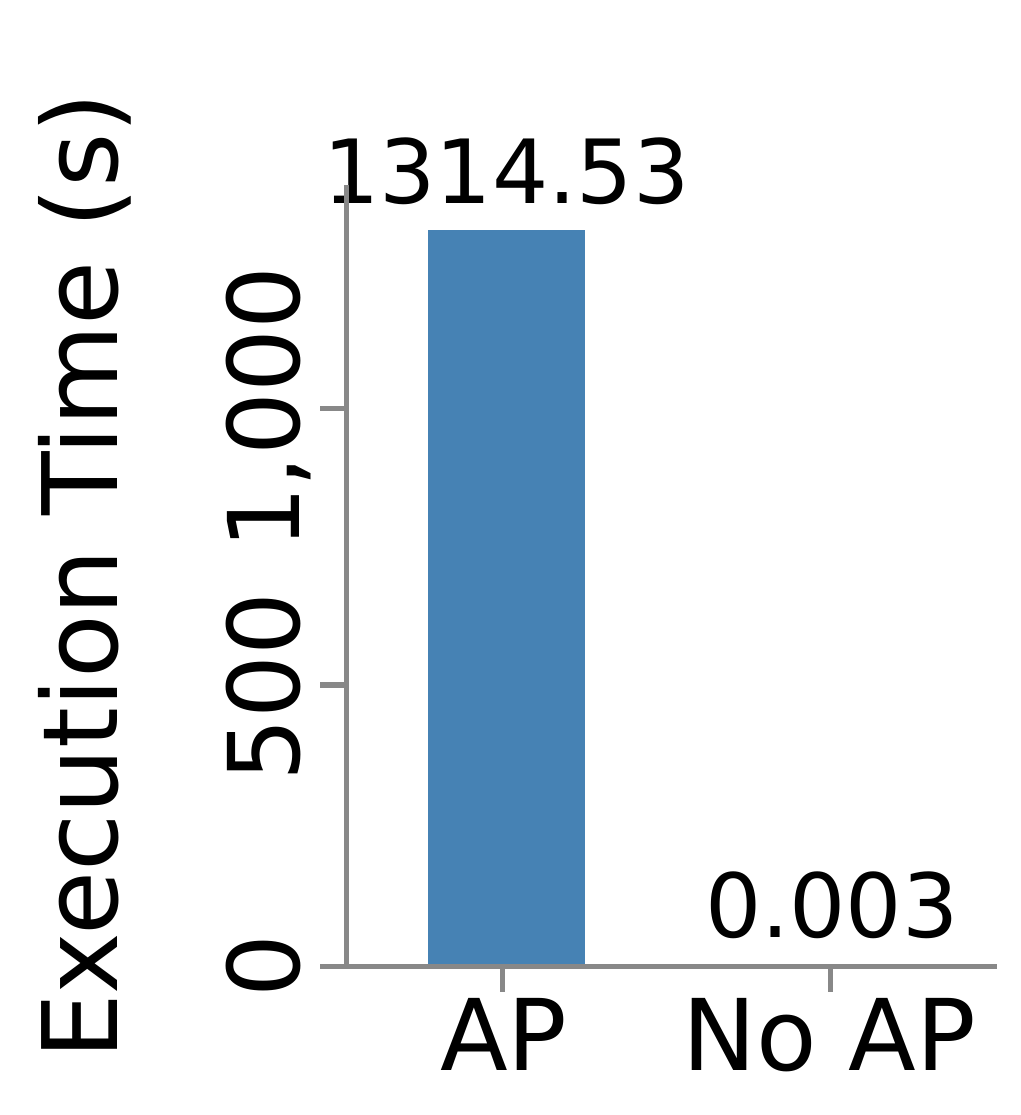}
\label{fig:value-in-definition-update}
}
\qquad
\subfloat[][Enumerated Types:\\ Insert]{
\centering
\includegraphics[width=0.12\textwidth]{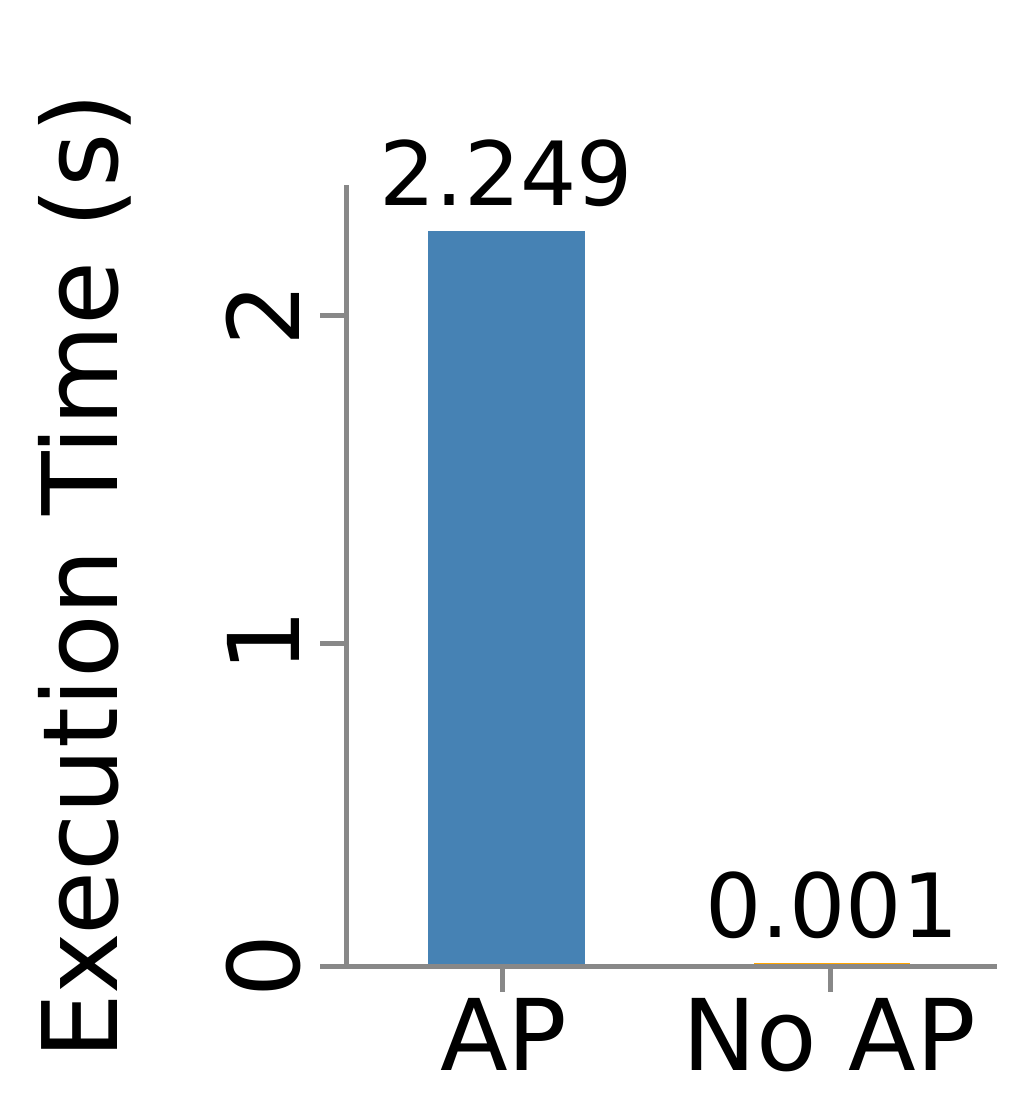}
\label{fig:value-in-definition-insert}
}
\qquad
\subfloat[][Enumerated Types:\\ Select]{
\centering
\includegraphics[width=0.12\textwidth]{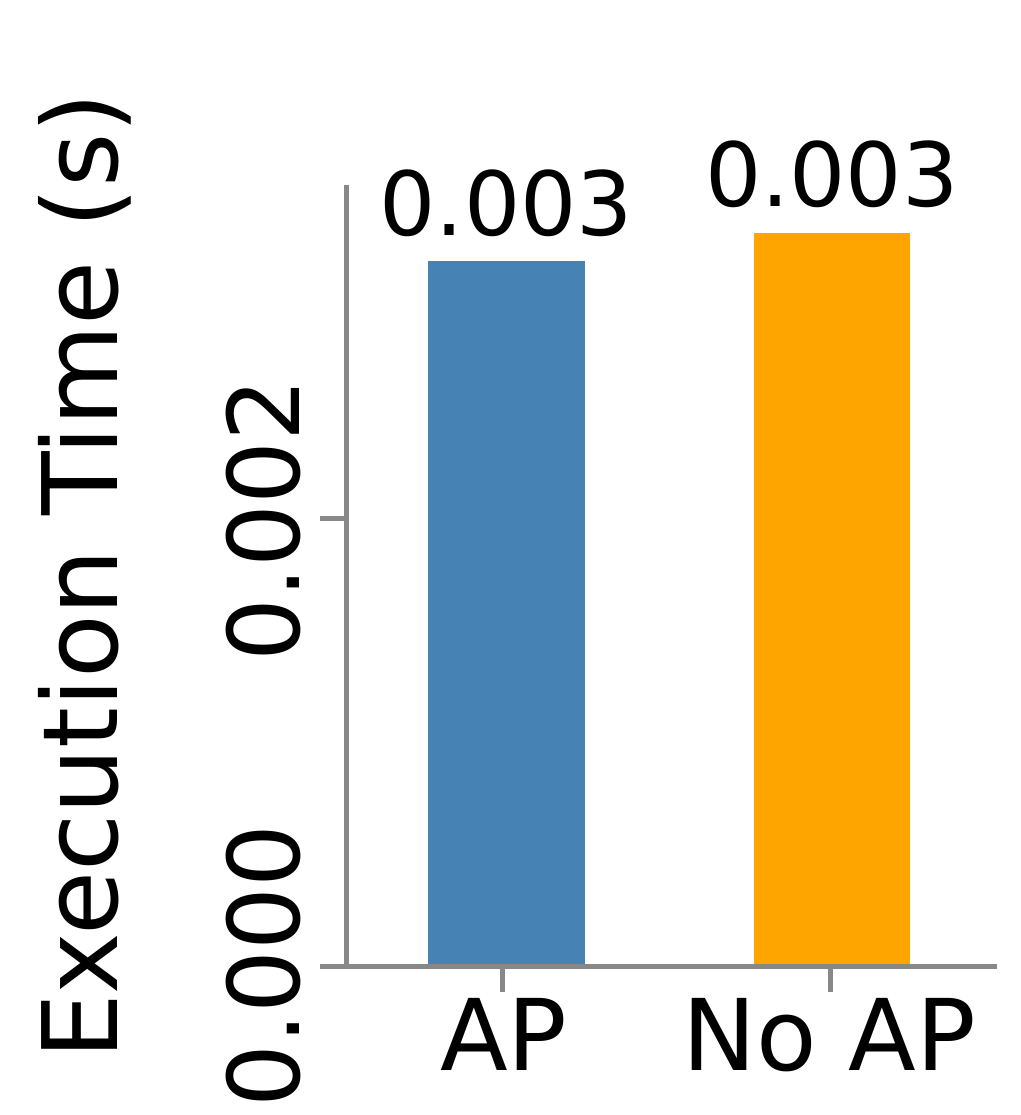}
\label{fig:value-in-definition-select}
}
\qquad
\caption{
\textbf{Ranking and Repair of \apshort:} Performance impact of \apshort on
different types of \sql statements.}
\end{figure*}


\PP{Index Overuse \apshort:}
This \apshort is associated with the creation of too many infrequently-used
indexes.
As shown in~\cref{fig:index-over-use-update}, the performance impact of this
\apshort is significant for the \texttt{UPDATE} statement.
We attribute this to the overhead of maintaining the indices. 
The update operation is 10 $\times$ slower when there are five indices on the
field being updated.
Thus, it is advisable to only create those indexes whose impact on query
processing is significant.
Another fix for this \apshort is to maintain a multi-column index as opposed to
maintaining multiple single-column indices.
%
%

%

\PP{Index Underuse \apshort:}
\cref{fig:index-underuse-select-aggr-wo-predicate} illustrates the impact of not having important
indices on columns.
We execute a query which performs a post-grouping aggregation operation. 
The query execution time drops by 1.3$\times$ when we create an 
index on the column contained in the \texttt{GROUP BY} clause. 
This is because the index eliminates the overhead of the grouping operation.
\cref{fig:index-underuse-select-predicate} illustrates a scenario where fixing
this \apshort reduces performance.
We consider a scan query with a predicate on a column with low cardinality.
The query executes 3$\times$ slower when it uses the index as opposed to a
table scan.
We attribute this to the low cardinality of the indexed
column~\cite{low-cardinality-index-problem}.
The query analysis rule incorrectly flags this query due to missing indices.
\sys eliminates this false positive by leveraging its data analysis rule which
takes cardinality into consideration.

\PP{No Foreign Key Exists \apshort:}
~\cref{fig:fk-exists-update,fig:fk-exists-select,fig:fk-exists-update-index}
illustrate the performance impact of this \apsshort on an \texttt{UPDATE}
statement and a scan query. 
The performance impact is not prominent in both cases 
since the PostgreSQL does \textit{not} automatically create an index to
maintain the foreign key constraint.
An index explicitly constructed by the user accelerates the \texttt{UPDATE}
operation by 142$\times$.
This \apshort also has a significant impact on maintainability and data
consistency.
This is because the foreign key constraint must be preserved by the developer
using complex application-level logic.
For instance, in case of a cascaded \texttt{DELETE} operation, a developer will
need to issue two \sql statements to update the values in both the reference and
referencing tables.
Otherwise, the referential integrity constraint  will not
be preserved.
This increases the complexity associated with maintaining the application.

\PP{Enumerated Types \apshort:}
With this \apshort, the developer uses the \texttt{CHECK} constraint feature of
the DBMS~\cite{pg-check}.
We measure the time taken to update a value of the \texttt{Role} column covered
by the constraint (\texttt{R2} $\mapsto$ \texttt{R5}).
This consists of three steps: (1) an \texttt{ALTER} operation to drop the
\texttt{CHECK} constraint on the \texttt{Role} column, 
(2) updating the value using an \texttt{UPDATE} statement based on a predicate,
and (3) an \texttt{ALTER} operation to add the constraint back onto the column.
In contrast, if the database contains an intersection table as shown
in~\cref{fig:pattern-1-good-3}, the same task can be accomplished using 
an \texttt{UPDATE} statement.
~\cref{fig:value-in-definition-update,fig:value-in-definition-insert,fig:value-in-definition-select}
illustrates the performance impact of this \apshort.
Eliminating this \apshort improves performance by more than 1000$\times$ in case
of \texttt{UPDATE} and \texttt{INSERT} operations.
The impact is less prominent in case of the scan query due to the
overhead of the \texttt{JOIN} operator.
The \apshort-free design improves maintainability by reducing the number of
queries required for performing a given task (\eg, \texttt{Role} update).
Furthermore, it reduces the storage footprint by reducing data duplication.

\PP{Severity of \apsshort:}
Impact of an \apshort depends on the application context (\eg, \ent \apshort will not affect  performance if the attribute's domain does not change).
Certain \apsshort may stem from application requirements.
For instance, it may not be possible to simplify a query with \tmj. 
Lastly, a subset of \apsshort will always have negative impact (\eg, \nfk will always affect data integrity).

\subsection{User Study}
\label{evaluation::userstudy}

In this experiment, we evaluate the efficacy of \sys through a user study.
We recruited 23 graduate and under-graduate students majoring in Computer
Science with varying degrees of expertise in \sql.

\PP{Experiment Setup:} 
We tasked the participants to construct a set of \sql queries for a bike
e-commerce application.
The requirements are twofold: (1) design a performant and extensible 
database design for this application, and  
(2) formulate performant \sql queries to support application features.
We curated a set of sixteen features that are associated with one or more
\apsshort (\eg, shopping cart, list of products).

The participants use the GUI Interface presented
in~\autoref{sec:implementation}.
%
We track: 
(1) the original \sql queries developed by the user, 
(2) the fixes suggested by \sys for the \apsshort detected in the original
queries, and 
(3) the re-formulated \sql queries developed by the user that incorporate these fixes.
We also collect qualitative feedback from the participants about the accuracy and utility of the detected \apsshort and their fixes.

\PP{Results:}
The participants constructed \texttt{987} \sql statements. 
\sys detected and suggested fixes for \texttt{207} \apsshort.
Most of the participants (\texttt{20} out of \texttt{23}) took the 
\texttt{187} \apsshort detected in their queries into consideration.
They refactored the queries to resolve \texttt{96} \apsshort.
They ignored the remaining \texttt{91} fixes.
The reasons for this are twofold: 
(1) ambiguous fixes (\texttt{31} fixes), and 
(2) incorrect fixes (\texttt{60} fixes) given the requirements of the application.
%
%
%
Thus, the participants leveraged 51\% of the fixes suggested by \sys. 
\new{If we also include the \apsshort labeled as ambiguous from the
participants' perspective, then the efficacy increases to 67\%.}


We compare the distribution of \apsshort detected in the participants' 
queries using \dbdeo and \sys. 
The results are shown
in~\cref{tab:ap-distribution-github-survey-dbdeo-sqlcheck}.
There is significant variation in the frequency of \apsshort.
For instance, the \npk \apshort is 14$\times$ more prevalent than the \pmr
\apshort.

We next examine the variance in \sql skills of the participants.
The number of queries executed, detected \apsshort, and accepted
\apsshort follow these distributions: ($\mu$=42.5, Q2=46),  
($\mu$=9.35, Q2=8), and ($\mu$=4.8, Q2=46). 
The high variance in \sql skills illustrates the need for an automated toolchain
for detecting \apsshort and suggesting fixes.
The qualitative feedback from the participants indicate that they predominantly
found \sys to be helpful in understanding \apsshort.

\begin{table}
    \centering
    \footnotesize
    \begin{tabular}{r | c c | c c | c c }
     & \multicolumn{2}{c|}{\textbf{GitHub Rep}} & \multicolumn{2}{c}{\textbf{User Study}} & \multicolumn{2}{c}{\textbf{\new{Kaggle}}} \\
     \textbf{Anti-Pattern} & \textbf{D} & \textbf{S} & \textbf{D} & \textbf{S} & \textbf{\new{D}} & \textbf{\new{S}} \\
\hline
No Primary Key & 628 & 6875 & 22 & 70 
& \new{-} & \new{68} \\
Column Wildcard Usage & 0 & 12313 & 0 & 54 
& \new{-} & \new{-} \\
Data in Metadata & 1907 & 1352 & 43 & 39 
& \new{-} & \new{9} \\
Enumerated Types & 90 & 462 & 11 & 30 
& \new{-} & \new{-} \\
Index Underuse & 82 & 506 & 40 & 30
& \new{-} & \new{-} \\
God Table & 3514 & 3371 & 22 & 28 
& \new{-} & \new{-} \\
Implicit Columns & 0 & 26488 & 0 & 24 
& \new{-} & \new{-} \\
Readable Password & 0 & 295 & 0 & 20 
& \new{-} & \new{-} \\
Clone Table & 1990 & 516 & 21 & 12 
& \new{-} & \new{-} \\
Rounding Errors & 1081 & 1426 & 91 & 10 & 
\new{-} & \new{-} \\
Generic Primary Key & 0 & 5123 & 0 & 8 & 
\new{-} & \new{25} \\
Multi-Valued Attribute & 2539 & 1503 & 3 & 6 & 
\new{-} & \new{20} \\
Pattern Matching & 552 & 1065 & 25 & 5 
& \new{-} & \new{-} \\
Adjacency List & 103 & 93 & 0 & 0 
& \new{-} & \new{-} \\
No Foreign Key & 0 & 1389 & 0 & 0 
& \new{-} & \new{10} \\
External Data Storage & 0 & 63 & 0 & 0 
& \new{-} & \new{-} \\
Index Overuse & 228 & 228 & 0 & 0 
& \new{-} & \new{-} \\
Concatenate Nulls & 0 & 63 & 0 & 0 
& \new{-} & \new{-} \\
Ordering by Rand & 0 & 27 & 0 & 0 
& \new{-} & \new{-} \\
Distinct and Join & 0 & 4 & 0 & 0 
& \new{-} & \new{-} \\
Too many Joins & 0 & 4 & 0 & 0 
& \new{-} & \new{-} \\
\new{Missing Timezone} & \new{-} & \new{-} & \new{-} & \new{-} & \new{-} & \new{12} \\
\new{Incorrect Data Type} & \new{-} & \new{-} & \new{-} & \new{-} & \new{-} & \new{28} \\
\new{Denormalized Table} & \new{-} & \new{-} & \new{-} & \new{-} & \new{-} & \new{16} \\
\new{Information Duplication} & \new{-} & \new{-} & \new{-} & \new{-} & \new{-} & \new{1} \\
\new{Redundant Column} & \new{-} & \new{-} & \new{-} & \new{-} & \new{-} & \new{11} \\
\new{No Domain Constraint} & \new{-} & \new{-} & \new{-} & \new{-} & \new{-} & \new{0} \\
\hline
\textbf{Total:} & 14764 & 63058 & 278 & 336 & \new{-} & \new{200} \\
\end{tabular}
    \caption{\textbf{Distribution of \apsshort --} 
    Distribution of \apsshort detected by \sys (S) and \dbdeo (D) 
    in queries collected from repositories on GitHub and written by the user study participants.
    }
    \label{tab:ap-distribution-github-survey-dbdeo-sqlcheck}
\end{table}

\subsection{Web Applications \& Databases}
\label{sec:evaluation:realworld-applications}

\new{
In this experiment, we first evaluate the efficacy of \sys in finding, ranking, and
suggesting fixes for \apsshort in real-world web applications on GitHub.
We apply \sys on \djangoappcount actively-developed Django-based
applications~\cite{django}.
}

\PP{Experiment Setup:} 
\new{
We first manually deploy each of these applications on PostgreSQL.
We then collect the \sql queries either by running the integration tests or by
manually interacting with the application.
Lastly, we report the high-impact \apsshort to the developers by either 
raising issues on GitHub or through the official developer forum.
Before reporting the \apsshort, we manually analyse them to study their 
significance based on the application-specific context.
We order the \apsshort impact metrics thus: read performance, maintainability,
write performance, accuracy, and amplification.}

\begin{table}[t]
    \centering
    \footnotesize
    \begin{tabular}{r l r r}
    \hline
    \textbf{\#} & 
    \textbf{GitHub Repo} & 
    \textbf{\# \apshort Det} &
    \textbf{\# \apshort Rep} \\
    \hline
    1
    & Globaleaks
    & 10
    & 2
    \\
    2
    & Django-oscar
    & 12
    & 2
    \\
    3
    & Saleor
    & 10
    & 2
    \\
    4
    & Django-crm
    & 8
    & 4
    \\
    5
    & django-cms
    & 11
    & 1
    \\
    % ...
    \hline
    \textbf{17} & \textbf{Total} & \textbf{123} & \textbf{32} \\ 
    \hline
    \end{tabular}
    \caption{\textbf{Evaluation of \sys on Web Applications:} 
    The \apsshort detected by \sys (\# APs Det) in
    a subset of \djangoappcount Django applications.
    We list the major \apsshort that we reported (\# APs Rep).
    }
    \label{tab:real-world-app-evaluation-short}
  \end{table}

\PP{Results:} 
As shown in~\cref{tab:real-world-app-evaluation-short}, \sys detected \djangoapcount
\apsshort across these applications (Ref. ~\cref{sec:appendix::real-world}).
We reported \djangomajorapcount \apsshort based on their impact score and the
application-specific context.
We do not report low severity \apsshort (\eg, \gpk) and those
that require a deeper understanding of application requirements (\eg, \tmj).

We have received responses from all but three of these development teams.
Eleven teams acknowledged the existence of \apsshort in their applications.
They attribute these \apsshort to the default behavior or lack of certain features in Django.
Four teams are incorporating the fixes from \sys.
Three teams are looking for alternate fixes.
Three teams did not share their course of action.
One team decided not to fix the reported \apsshort given their
application-specific requirements.
Most of these teams were interested in understanding the implications of these
\apsshort and requested us to send patches.
In one of these applications, \sys found \apsshort that introduced by a
third-party library.
In another application, we found an existing issue related to the \tmj \apshort.
The developers found that replacing the ORM-generated query with a simpler,
hand-written query greatly improved performance.
This experiment illustrates the efficacy of \sys in assisting application
developers in practice.
%


\PP{Data Analysis:}
\new{
We next evaluate the efficacy of \sys in finding \apsshort in real-world databases on Kaggle~\cite{kaggle}.
We download \kaggledbcount SQLite databases and apply the data analysis rules of \sys on them (\autoref{sec:finding::data}). 
As shown in~\cref{tab:real-world-dataset-evaluation-short}, \sys detects
\kaggleapcount \apsshort across these databases
(Ref. \cref{sec:appendix::data-analysis}).
}
\new{This experiment illustrates the efficacy of \sys in detecting \apsshort by only analysing data (without queries).}

\begin{table}[t]
    \centering
    \footnotesize
    \begin{tabular}{r l r}
    \hline
     \textbf{\#} & \textbf{Kaggle Database} & \textbf{\# AP}
     \\
     \hline
     1 & The History of Baseball & 41 \\
     2 & Soccer Dataset & 20 \\
     3 & Acad. Research from Indian Univ. & 17 \\
     4 & Pesticide Data Program & 13 \\
     5 & Board Games & 12 \\
    \hline
    \textbf{31} & \textbf{Total} & \textbf{200}\\
 \hline
 \end{tabular}
    \caption{\textbf{Evaluation of \sys on real-world databases:} 
    The \apsshort detected by \sys in a subset of \kaggledbcount Kaggle databases.}
    \label{tab:real-world-dataset-evaluation-short}
  \end{table}
\subsection{Limitations And Future Work}
\label{sec:eval::limitations}


\PP{Anti-Pattern Coverage, Discovery, and Evolution:}
\sys currently detects \new{\totalapsupportcount} types of \apsshort. 
We intend to add support for more known \apsshort in the future.
However, it is unclear how to automatically discover new types of \apsshort in
\sql queries.
Furthermore, the performance impact of an \apshort can evolve over time.
For instance, the performance impact of the \textit{Adjacency list} \apshort was
prominent in PostgreSQL v9 (5$\times$).
However, it is no longer significant (1.1$\times$) in v11.

\PP{Dialect Coverage and Query Repair:}
\sys is designed to support multiple \sql dialects for higher utility.
We accomplish this using a non-validating query parser
(\autoref{sec:finding::query}).
However, it is infeasible to handle dialect-specific features, especially in
complex queries.
The usage of a non-validating query parser also restricts the set of queries
wherein we can automatically rewrite the query to fix the \apshort.
This is because we do not have enough syntactical information for query
rewriting.
We instead fall back on tailored textual fixes in these scenarios.
We made this decision to increase the utility of \sys.
The data analyzer (\autoref{sec:finding::data}) is built on top of
\textsc{SQLAlchemy} so that it can support diverse DBMSs (\eg, PostgreSQL,
MySQL).
Thus, the set of DBMSs that can be analyzed using \detector is constrained by
those that are supported by \textsc{SQLAlchemy}.

%

\section{Related Work}{\label{sec:background}}

\PP{Transforming Database Applications:}
Although program analysis has a long history in software
engineering, it has not been extensively studied by the DBMS community.
Recent research efforts have focused on transforming database-backed
programs to improve performance~\cite{cheung14,yan16,guravannavar08}. 
Ramachandra \etal present application transformations that enable asynchronous
and batched query submission~\cite{ramachandra15}.
DBridge presents a set of holistic optimizations including query batching and
binding, and automatic transformation of object-oriented code into
synthesized queries~\cite{emani2016,ramachandra15}.
Cheung \etal describe techniques for 
batching queries to reduce the number of round trips between the application 
and database servers~\cite{cheung13,cheung16-b,cheung14,cheung13-b}.


\PP{Object-Relational Mapping:}
Researchers have studied the impact of ORM on application design and
performance~\cite{torres17,chen14,yan17,chen15,chen16}. Yang \etal perform a
comprehensive study of performance issues in database applications using profiling
techniques~\cite{yang18}. 

This paper is the first to explore the problems of automatically ranking and
fixing \apsshort in database applications. 

\section{Conclusion}
{\label{sec:conclusion}}
In this paper, we presented \sys, a holistic toolchain for finding, ranking, 
and fixing \apsshort in database applications.
\sys leverages a novel \apshort detection algorithm that augments 
query analysis with data analysis.
It improves upon \dbdeo, the state-of-the-art tool for detecting \apshort,
by using the overall context of the application to reduce false positives and
negatives.
\sys relies on a ranking model for characterizing the impact of detected 
\apsshort.
%
%
We discussed how \sys suggests fixes for high-impact \apshort 
using rule-based query refactoring techniques.
%
%
Our empirical analysis shows that \sys enables developers to create
more performant, maintainable, and accurate applications.

\vspace{2pt}
\noindent \textbf{\large Acknowledgements}\\
{ 
This work was supported in part by the U.S. National Science
Foundation
(\href{https://www.nsf.gov/awardsearch/showAward?AWD_ID=1908984}{IIS-1908984}, \href{https://www.nsf.gov/awardsearch/showAward?AWD_ID=1850342}{IIS-1850342}),
Intel, and Alibaba.
We thank Shamkant Navathe and our reviewers for their constructive feedback. 
We thank all of the contributors to \sys: 
Venkata Kishore Patcha, Varsha Achar, Pooja Bhandary, Jennifer Ma, and 
Sri Vivek Vanga.}

\newpage
{
\bibliographystyle{ACM-Reference-Format}
\small
\raggedright
\bibliography{sqlcheck}
}

\null\newpage
\begin{appendices}
\section{Data Analysis}
\label{sec:appendix::data-analysis}

In this experiment, we apply \sys's data analysis rules on \kaggledbcount
publicly-available SQLite databases from Kaggle.
\cref{tab:data-analysis-results} lists the SQLite databases from
Kaggle~\cite{kaggle} that we use in this experiment along with  the \apsshort
detected in these databases. 
We found \kaggleapcount \apsshort across \kaggledbcount databases using \sys.
The results of this experiment are discussed in
~\autoref{sec:evaluation:realworld-applications}. 

\begin{table*}[t]
    \centering
    \footnotesize
    \begin{tabular}{r l r l}
   \hline
    \textbf{\#} & \textbf{SQLite Database} & \textbf{\# AP} &
    \textbf{Detected Anti-Patterns}
    \\
    \hline
    1 & Board Games & 12 & No Primary Key, Data in Metadata, Incorrect Datatype \\
    2 & Pennsylvania Safe Schools Report & 1 & No Primary Key \\
    3 & Soccer Dataset & 20 & Generic Primary Key, Data in Metadata, Missing Timezone, Multivalued Attribute \\
    4 & SF Bay Area Bike Share & 11 & No \& Generic Primary Key, Incorrect Datatype, Missing Timezone, Denormalized Table \\
    5 & US Baby Names & 2 & Generic Primary Key \\
    6 & Pitchfork Music Data & 10 & No Primary Key, Missing Timezone, Information Duplication, Denormalized Table \\
    7 & Acad. Research from Indian Univ. & 17 & No Primary Key, Incorrect Datatype, Redundant Column, Multivalued Attribute \\
    8 & What.CD HipHop & 3 & No Primary Key, Multivalued Attribute \\
    9 & Snap Meme-Tracker & 1 & Missing Timezone \\
    10 & NIPS papers & 4 & Generic Primary Key, Denormalized Table \\
    11 & US Wildfires & 2 & No Primary Key, Redundant Column \\
    12 &  Que from crossvalidated StackExc & 3 & No Primary Key \\
    13 & The History of Baseball & 41 & No Primary Key, Data in Metadata, Incorrect Datatype, Multivalued Attribute \\
    14 & Twitter US Airline Sentiment & 2 & Denormalized Table \\
    15 & Hilary Clinton Emails & 8 & Generic Primary Key, Incorrect Datatype \\
    16 & SEPTA - Regional Rail & 2 & Incorrect Datatype, Missing Timezone \\
    17 & US Consumer finance Complaints & 9 & No Primary Key, Incorrect Datatype, Multivalued Attribute, Denormalized Table \\
    18 & 1st GOP Debate Twitter Sentiment & 1 & Generic Primary Key \\
    19 & SF Salaries & 2 & Generic Primary Key, Denormalized Table \\
    20 & Freight Matrix Transportation & 5 & No Primary Key, Data in Metadata, Redundant Column \\
    21 & WDIdata & 9 & No Primary Key, Multivalued Attribute \\
    22 & Amazon Movie Reviews Dataset & 2 & No Primary Key, Multivalued Attribute \\
    23 & UK Arms Export License & 3 & No Primary Key \\
    24 & Amazon Fine Food Reviews & 1 & Generic Primary Key \\
    25 & Stackoverflow Question Favourites & 1 & Multivalued Attribute \\
    26 & Iron March & 1 & Redundant Column \\
    27 & C\# Methods with Doc. Comments & 4 & Generic Primary Key \\
    28 & Pesticide Data Program & 13 & No Primary Key, Incorrect Datatype, Redundant Column \\
    29 & Monty Python Flying Circus & 4 & No Primary Key, Missing Timezone, Denormalized Table \\
    30 & Twitter Conv. about Black Panther & 0 & - \\
    31 & 2016 US Election & 6 & No Primary Key, Data in Metadata, Denormalized Table \\
    \hline
    & \textbf{Total} & \textbf{200} & \\
\hline
\end{tabular}
    \caption{\textbf{Data Analysis Results}
    A list of SQLite databases from Kaggle ~\cite{kaggle} with the 
    \apsshort detected by applying the data analysis rules in \sys.
    }    
    \label{tab:data-analysis-results}
\end{table*}

\begin{table*}[t]
  \centering 
  \footnotesize
  
\newcolumntype{Y}{>{\centering\arraybackslash}X}
\begin{tabularx}{\textwidth}{@{}lrrlrlll@{}}
\toprule
\textbf{GitHub Repo} & 
\textbf{Stars} &
\textbf{Contr.} &  
\textbf{Domain} &
\textbf{\# \apshort} &
\textbf{\apsshort Reported} &
\textbf{R} &
\textbf{A} \\

\midrule
\multirow{1}{*}{Globaleaks}
& 741
& 22
& Whistleblower
& 10
& 2 (No Foreign Key, Enumerated Types)
& $\checkmark$  
& $\checkmark$ 
\\
\multirow{1}{*}{Django-oscar}
& 4.1k
& 217
& E-commerce 
& 12
& 2 (Rounding Errors, Index Overuse)
& $\checkmark$  
& $\checkmark$ 
\\
\multirow{1}{*}{Saleor}
& 6.5k
& 139
& E-commerce 
& 10
& 2 (Multivalued Attribute, Index Overuse)
& $\checkmark$  
& $\checkmark$ 
\\
\multirow{1}{*}{Django-crm}
& 654
& 17
& CRM
& 8
& 4 (Index Underuse, Index Overuse, Pattern Matching, No Domain Constraint)
& $\checkmark$  
& $\checkmark$ 
\\
\multirow{1}{*}{django-cms}
& 7.2k
& 398
& CMS
& 11
& 1 (Index Overuse)
& $\checkmark$  
& $\checkmark$ 
\\
\multirow{1}{*}{wagtail-autocomplete}
& 41
& 7
& Utility 
& 1
& 1 (Pattern Matching)
& $\checkmark$  
& $\checkmark$ 
\\
\multirow{1}{*}{shuup}
& 1.1k
& 41
& E-commerce 
& 6
& 1 (Index Overuse)
& $\checkmark$  
& $\checkmark$ 
\\
\multirow{1}{*}{Pretix}
& 821
& 113
& E-commerce 
& 11
& 3 (Index Overuse, Pattern Matching, No Domain Constraint)
& $\checkmark$  
& $\checkmark$ 
\\
\multirow{1}{*}{Django-countries}
& 755
& 35
& Library 
& 1
& 1 (Multivalued Attribute)
& $\checkmark$  
& $\checkmark$ 
\\
\multirow{1}{*}{micro-finance}
& 55
& 8
& Finance 
& 8
& 4 (Index Underuse, Index Overuse, Pattern Matching, No Domain Constraint)
& $\checkmark$  
& $\checkmark$ 
\\
\multirow{1}{*}{bootcamp}
& 1.9k
& 24
& Social Ntwrk 
& 5
& 1 (Index Overuse)
& $\checkmark$  
& $\checkmark$ 
\\
\multirow{1}{*}{NetBox}
& 6.2k
& 118
& DCIM
& 9
& 3 (Index Overuse, Pattern Matching, No Domain Constraint)
& $\checkmark$  
& $\checkmark$ 
\\
\multirow{1}{*}{Ralph}
& 1.3k
& 43
& Asset Mgmt
& 12
& 3 (Index Overuse, Pattern Matching, No Domain Constraint)
& $\checkmark$  
& $\times$
\\
\multirow{1}{*}{Tiaga}
& 6.5k
& 139
& E-commerce 
& 9
& 2 (Index Overuse, No Domain Constraint)
& $\checkmark$  
& $\times$ 
\\
\multirow{1}{*}{wagtail}
& 8.4k
& 397
& CMS 
& 10
& 2 (Index Overuse, No Domain Constraint)
& $\checkmark$
& $\times$ 
\\
\hline
\textbf{Total} &  &  &  & \textbf{123} & \textbf{32} & & \\
\bottomrule
\end{tabularx}
  \caption{\textbf{Evaluation of \sys on real-world applications on GitHub:} 
  A list of open-source database applications on GitHub along with their 
  popularity and domain, the \apsshort (counts and names) detected by \sys, 
  whether these \apsshort have been reported \textbf{(R)} as GitHub Issues 
  and acknowledged \textbf{(A)} by the contributors.
  }
  \label{tab:real-world-app-evaluation}
\end{table*}

\section{Real World Applications}
\label{sec:appendix::real-world}

We tested \sys on \djangoappcount actively developed, open-source applications
based on the Django ORM~\cite{django}.
In this experiment, we pass the \sql queries obtained from either running the
application's integration tests or by manually interacting with the application 
to \sys.
We rank the detected \apsshort based on their impact and reported the ones with
high impact. The \apsshort detected are summarized
in~\cref{tab:real-world-app-evaluation}.
The results of this experiment are discussed in
~\autoref{sec:evaluation:realworld-applications}. 

\section{Database Tuning Advisors}

There is a large body of work related to automated database administration tools
geared towards optimizing: (1) database physical design,
and (2) database knob configuration.
The former set of tools focus on selecting the best physical design (\eg,
materialized views) for a given database that
optimizes the target metrics (\eg, latency) and satisfies the budget constraints
(\eg, hardware resources).
These tools include Microsoft's AutoAdmin~\cite{chaudhuri98} and
DETA~\cite{deta}
The latter set of tools instead focus on  selecting the best knob configuration
(\eg, buffer pool size) for a given DBMS.
These tools include OtterTune~\cite{ottertune} and
vendor-specific offerings~\cite{dias05,kumar03,narayanan05,ibmdb02}.

\begin{table*}[t]
    \centering
    \footnotesize
    \begin{tabular}{l c c }
   \hline
    \textbf{Supported Features} & \textbf{DETA}  & \textbf{SQLCheck}
    \\
    \hline
    Index creation/destruction suggestions & $\checkmark$ & $\checkmark$
    \\
    Type of index to create based on workload  & $\checkmark$ & $\times$ \\
    Materialized view creation/destruction suggestions &
    $\checkmark$ & $\times$ \\
    Suggestions tailored based on hardware constraints, workload, \& data
    distribution  & $\checkmark$ & $\times$ \\
    Table partitioning suggestions & $\checkmark$ & $\times$
    \\
    Column type suggestions based on data & $\times$ & $\checkmark$ \\
    Query refactoring suggestions & $\times$ & $\checkmark$ \\
    Alternate logical schema design suggestions & $\times$ & $\checkmark$
    \\
    Logical errors that may invalidate data integrity & $\times$ & $\checkmark$ \\
\hline
\end{tabular}
    \caption{\textbf{SQLCheck v/s Microsoft DETA~\cite{deta}}
    Comparison of the core features of \sys against a physical design tuning
    advisor.
    }    
    \label{tab:sqlcheck-vs-deta}
\end{table*}

\cref{tab:sqlcheck-vs-deta} compares the core features of \sys and a tuning advisor (DETA).
\sys complements DETA in that it focuses on other aspects of database
applications besides physical design (\eg, logical design \apsshort and query
\apsshort).
We recommend the user to leverage the accurate feedback of the physical design
tuning tool to fix the detected physical design \apsshort, since it leverages
the cost model of the query optimizer.
\sys assists application developers in preventing \apsshort during the
application development phase by suggesting better alternatives 
(as opposed to tuning tools that are more often used post deployment).  

\end{appendices}

\end{document}